\documentclass[fleqn,usenatbib]{mnras}

\usepackage{newtxtext,newtxmath}

\usepackage[T1]{fontenc}

\DeclareRobustCommand{\VAN}[3]{#2}
\let\VANthebibliography\thebibliography
\def\thebibliography{\DeclareRobustCommand{\VAN}[3]{##3}\VANthebibliography}

\usepackage{graphicx}	
\usepackage{amsmath}	
\usepackage{mathtools}
\usepackage{mhchem}
\usepackage{gensymb}
\usepackage{hyperref}

\title[Equatorial Jets]{Longitudinal Filtering, Sponge Layers, and Equatorial Jet Formation in a General Circulation Model of Gaseous Exoplanets}

\author[D. A. Christie et al.]{D. A. Christie,$^{1,2}$\thanks{E-mail: christie@mpia.de}
N. J. Mayne,$^{2}$
M. Zamyatina,$^{2}$
H. Baskett,$^{2}$
T. M. Evans-Soma,$^{1,3}$
N. Wood,$^{4}$\newauthor
and K. Kohary$^{2}$
\\
$^{1}$Max Planck Institute for Astronomy, K\"onigstuhl 17, D-69117 Heidelberg, Germany\\
$^{2}$Department of Physics and Astronomy, Faculty of Environment, Science and Economy, University of Exeter, Exeter EX4 4QL, UK\\
$^{3}$School of Information and Physical Sciences, University of Newcastle, Callaghan, NSW, 2308, Australia\\
$^{4}$Met Office, Fitzroy Road, Exeter EX1 3PB, UK
}

\date{Accepted XXX. Received YYY; in original form ZZZ}

\pubyear{2024}

\begin{document}
\label{firstpage}
\pagerange{\pageref{firstpage}--\pageref{lastpage}}
\maketitle

\begin{abstract}
General circulation models are a useful tool in understanding the three dimensional structure of hot Jupiter and sub-Neptune atmospheres; however, understanding the validity of the results from these simulations requires an understanding the artificial dissipation required for numerical stability.  In this paper, we investigate the impact of the longitudinal filter and vertical ``sponge'' used in the Met Office's {\sc Unified Model} when simulating gaseous exoplanets.   We demonstrate that excessive dissipation can  result in counter-rotating jets and a catastrophic failure to conserve angular momentum.  Once the dissipation is reduced to a level where a super-rotating jet forms, however, the jet and thermal structure are relatively insensitive to the dissipation, except in the nightside gyres where temperatures can vary by $\sim 100\,\mathrm{K}$.  We do find, however, that flattening the latitudinal profile of the longitudinal filtering alters the results more than a reduction in the strength of the filtering itself.   We also show that even in situations where the temperatures are relatively insensitive to the dissipation, the vertical velocities can still vary with the dissipation, potentially impacting physical processes that depend on the local vertical transport.  
\end{abstract}




\section{Introduction}

General circulation models (GCMs) are an important tool in understanding the three-dimensional nature of hot Jupiter atmospheres given their potential for fast winds and large day-night temperature gradients.  GCMs are faced, however, with the issue that simulations often require the inclusion of artificial dissipation to suppress grid-scale fluctuations in order to maintain numerical stability, potentially altering the results, limiting our ability to reproduce observations, and taking the simulations in unphysical directions.   Furthermore, there are physical dissipation mechanisms within hot Jupiter atmospheres that are ignored \citep[e.g., gravity waves, shear instabilities;][]{watkins_2010,fromang_2016,carone_2020} or only approximated \citep[e.g., magnetic drag;][]{perna_2010a,rauscher_2013,beltz_2022} leading to a situation where the inclusion of a parametrised dissipation mechanism, representing the entirety of the unmodelled physics, may well be required to reproduce observed wind speeds even in high resolution simulations, as discussed in \citet{heng_2011} and \citet{li_2010}.

Understanding the intrinsic dissipation in our models and the impact of dissipation included for numerical stability therefore becomes important as we seek to disentangle the dissipation required by unmodelled physics from the dissipation added due to numerical requirements.  Intercomparisons of GCMs comparing idealised test cases, such as \citet{polichtchouk_2014}, can provide robust tests of individual components of a model while intercomparisons of specific planetary targets, such as THAI \citep{turbet_2022,sergeev_2022,fauchez_2022}, CAMEMBERT \citep{christie_2022b},  and MOCHA (Iro et al., in prep), can provide insights about the model variability across the range of GCMs in the presence of many coupled physical processes.  These sorts of investigations, however, are often done for fixed, specific choices of the dissipation, as their intention is to look for differences between models, not within the models themselves.  Some investigations of dissipation with these models do exist, however.  The impact of  dissipation order and effective resolution was investigated by \citet{skinner_2021} within their pseudospectral code, concluding that an effective grid of $\sim 2000\times 1000$ with a $\nabla^{16}$ diffusion operator is required to achieve numerical convergence within their model. \citet{heng_2011} also investigated the role of dissipation as a part of a larger investigation of multiple dynamical cores used within the dual-grey model Flexible Modelling System ({\sc fms}) GCM. They found that variation between solution method and level of dissipation resulted in uncertainties in the wind speeds on the order of tens of percent, concluding that observational data are required to constrain the appropriate level of dissipation.  While not varying the dissipation required for numerical stability, \citet{koll_2018} did investigate the role of explicit drag, approximating unmodelled physics, in the determination of the jet speed and examined the transition from Rayleigh drag dominating the dissipation of kinetic energy to the regime where numerical dissipation dominates.   \citet{hammond_2022} examined the stabilising hyperdiffusion in the {\sc thor} GCM and found that changes in diffusion order can result in zonal velocities changing by 50\% with lower-order hyperdiffusion resulting in a slower equatorial jet.

In this paper, we examine the role of the longitudinal filter and the vertical sponge used within Met Office's {\sc Unified Model} (UM) to stabilize simulations of hot Jupiters and sub-Neptunes, using a model of a $10\times$ solar metallicity WASP-96b previously investigated by \citet{zamyatina_2024} as a test case.  This paper is structured as follows:  in Section \ref{Sec:Numerics}, we outline the numerical setup used with the results presented in Section \ref{Sec:Results}.  A summary of the conclusions is presented in Section \ref{Sec:Conclusions}. In Appendix \ref{Apdx:Eddy}, we present the eddy momentum flux gradients for each of the simulations presented in the paper, and in Appendix \ref{Apdx:Res}, we present results from a short simulation with an increased horizontal resolution.

\section{Numerical Setup}
\label{Sec:Numerics}
In this section we outline the details of the Met Office's {\sc unified model} (UM) GCM as used in simulations of gaseous exoplanets, the details of the modeled planet WASP-96b, and the outline of the parameter studies that have been conducted.

\subsection{The {\sc Unified Model}}

The UM, through its {\sc Endgame} dynamical core, solves the full, deep-atmosphere, non-hydrostatic Navier Stokes equations \citep{mayne_2014a,wood_2014} and has been adapted to model hot Jupiters \citep[e.g.,][]{mayne_2014b,amundsen_2016} and sub-Neptunes \citep[e.g.,][]{drummond_2018,mayne_2019}. The radiative transfer is handled by the {\sc Socrates} radiative transfer code based on \citet{edwards_1996} which was adapted to the \ce{H_2} and \ce{He} dominated atmosphere appropriate to hot Jupiters and sub-Neptunes in \citet{amundsen_2014a,amundsen_2016,amundsen_2017}.  Within the configuration adopted here, previously used in \citet{christie_2021}, the included opacity sources are \ce{H_2O}, \ce{CO}, \ce{CH_4}, \ce{NH_3}, \ce{Li}, \ce{Na}, \ce{K}, \ce{Rb}, \ce{Cs} and \ce{H_2}-\ce{H_2} and \ce{H_2}-\ce{He} collision-induced absorption (CIA). Opacities are computed using the correlated-k method and {\sc ExoMol} line lists \citep{tennyson_2012,tennyson_2016}, and the chemical abundances are computed using the analytic model of \citet{burrows_1999} which was extended by \citet{amundsen_2016} to include alkali species.   While the {\sc UM} is capable of more sophisticated chemical modelling \citep[e.g.,][]{drummond_2018,zamyatina_2023,zamyatina_2024}, as the purpose here is to investigate the impact of dissipation choices, the analytic model is sufficient.

To achieve stability, the {\sc UM} implements two methods of dissipation and diffusion in the hot Jupiter/sub-Neptune setup.   To limit reflections at the upper computational boundary, a vertical ``sponge'' is employed which damps vertical velocity $w$ near the boundary, 

\begin{equation}
    \left(\frac{\partial w}{\partial t}\right)_\mathrm{sponge} =  -k_\mathrm{sp} w\,\, .
\end{equation}

\noindent The height-dependent dissipation coefficient $k_\mathrm{sp}$ is given by

\begin{equation}
k_\mathrm{sp} = \begin{cases} 
C\sin\left(\frac{\pi(\eta-\eta_s)}{2(1 - \eta_\mathrm{s})}\right) & \eta \geq \eta_s \\
0 & \eta < \eta_s
\end{cases}
\label{Eqn:Sponge}
\end{equation}

\noindent where $z$ is the vertical coordinate with $z=0$ at the base of the computational domain, $\eta = z/z_\mathrm{top}$, $\eta_s$ is the dimensionless height at which the sponge layer begins relative to the top of the top of the computational domain $z_\mathrm{top}$, and $C$ is a coefficient regulating the strength of the vertical sponge\footnote{\citet{drummond_2018} implemented a polar profile to the sponge wherein the vertical coordinate $\eta$ used in determining $k_\mathrm{sp}$ has a latitudinal dependence,

\begin{equation}
\eta(\phi) = \frac{z}{z_\mathrm{top}}\cos\phi + (1 - \cos\phi)\, .
\end{equation}

\noindent where $\phi$ is the latitude with the equator located at $\phi=0$.  This change has the effect of causing the sponge layer to descend to lower altitudes near the poles, with $\eta_s$ only representing the height at which the sponge layer begins at the equator.   Both the horizontally uniform and the polar sponge profile have been used in simulations of hot Jupiters and mini-Neptunes.}.  As the height at which the sponge begins is measured relative to the location of the upper boundary, the effect of changing $\eta_s$ can be accomplished by raising or lowering the upper boundary, subject to issues of numerical stability.   The use of a sponge to limit reflections at the upper boundary is common in studies of both planetary and exoplanetary atmospheres, although the details vary between specific implementations \citep[see e.g.,][]{carone_2020, deitrick_thor_2020}.

The second form of dissipation is a longitudinal filtering of the horizontal wind ${\bf u}_\mathrm{h} = (u,v)$,
\begin{equation}
    \left({\bf u}_\mathrm{h}\right)^{n+1}_{i,j,k}  = \left({\bf u}_\mathrm{h}\right)^n_{i,j,k} + K\left[ \left({\bf u}_\mathrm{h}\right)^n_{i+1,j,k} - 2 \left({\bf u}_\mathrm{h}\right)^n_{i,j,k} +  \left({\bf u}_\mathrm{h}\right)^n_{i-1,j,k}\right]\,\, , 
\end{equation}

\noindent where $i$, $j$, and $k$ index the longitudinal, latitudinal, and vertical zones, respectively, and the filtering constant $K$ is given by

\begin{equation}
    K = \frac{1}{4}\left(1-e^{-\frac{1}{t_K}}\right)\,\,,
\end{equation}
\noindent and is parametrised by $t_K$ which is the number of applications of the operator needed for a grid-scale perturbation to decay by a factor of $e$.  The first order Shapiro filter corresponds to the limit of $t_K\rightarrow 0$, or, equivalently, $K=1/4$.  For many hot Jupiter simulations performed using the UM \citep[e.g.,][]{mayne_2017,lines_2018a,christie_2021}, the value of $t_K$ is taken to be $1$; however, it has been found that this level of dissipation can, in certain cases greatly influence the dynamics \citep[e.g.,][]{christie_2022,zamyatina_2024}.   This choice of filtering is equivalent to a longitudinal diffusion operator applied to the horizontal wind,

\begin{equation}
    \left(\frac{\partial \bf{u}_\mathrm{h}}{\partial t}\right)_\mathrm{polar\,filter} = K_\mathrm{eff}\nabla^2_\lambda \bf{u}_\mathrm{h}\,\, ,
\end{equation}

\noindent where $\lambda$ is the longitude.  The effective longitudinal diffusion coefficient $K_\mathrm{eff}$ is given by
\begin{align}
    K_{\mathrm{eff}} & \sim K\frac{r^2\cos^2\phi\left(\Delta\lambda\right)^2}{\Delta t_\mathrm{dyn}}\,\, ,
    \label{Eqn:Keff}
\end{align}

\noindent where $\phi$ is the latitude with $\phi=0\degree$ being the equator, $\Delta \lambda$ is the longitudinal grid spacing, and $\Delta t_\mathrm{dyn}$ is the dynamical timestep.

We note that as this filter operates exclusively on the horizontal velocity without a corresponding transport of gas density, neither the linear nor the angular momentum are conserved throughout this process, except in the exceptional case where there are not longitudinal gradients in gas density. While applying the filter instead to the horizontal momentum $\rho{\bf u}_h$ would improve conservation of axial angular momentum, it would not necessarily achieve the desired result of suppressing grid-scale perturbations in horizontal velocities as only perturbations in momentum would be filtered.  Higher order approaches can avoid this issue \citep[e.g.,][]{deitrick_thor_2020}; however, this is beyond the scope of this paper.

This choice of stabilizing dissipation differs from other GCMs in that it is a lower-order operator and applied only to the horizontal winds, and only in the longitudinal direction.    While the focus of the operator is to suppress grid-scale perturbations and noise, it also has the effect of introducing a latitudinally-dependent diffusion acting on the horizontal winds due to the filter acting on a latitude-longitude grid.  

Unlike other studies \citep[e.g.,][]{liu_2013, carone_2020,schneider_2022}, studies using the UM have not included a ``Rayleigh'' drag at the inner boundary. As with the other types of dissipation, this paramtrisation can be included purely for reasons of numerical stability \citep{carone_2020,schneider_2022}, or it could represent a real physical mechanism such as  magnetic drag \citep[e.g,,][]{liu_2013}, although these parametrisations remain poorly constrained. Studies employing this interior drag could remove spurious accelerations in the deep atmosphere, but could also artificially impact the angular momentum exchange between the deep and upper atmosphere (e.g., \citealt{liu_2013, carone_2020}, also see \citealt{schneider_2022}, Appendix A). 

\subsection{WASP-96b}

For the purposes of investigating the impact of the longitudinal filtering and damping on the dynamics, we opt to model the atmosphere of WASP-96b.  While the polar filtering can be seen to impact models of other targets resulting in counter-rotating jets, such as with the GJ~1214b $100\times$ solar metallicity, $f_\mathrm{sed}=0.1$  case in \citet{christie_2022}, WASP-96b was seen to show this behaviour \citep[see ][]{zamyatina_2024} that can be easily reproduced without requiring additional computationally-intensive physics (clouds, non-equilibrium chemistry, etc.).

WASP-96b's metallicity is taken to be $10\times$ solar metallicity, with the abundances of all species other than \ce{H_2} and \ce{He} having their abundances relative to \ce{H} increased by a factor of $10$, with the elemental abundances taken from \citet{lodders_2009}.  The analytic chemistry formulation of \citet{burrows_1999} is found to be sufficient to reproduce the issue of counter-rotating jets and a breakdown in the conservation of angular momentum, so while \citet{zamyatina_2024} employs equilibrium and non-equilibrium chemistry schemes, we deem those unnecessary for this study and opt for computational simplicity.

The simulation parameters are presented in Table \ref{Tbl:Common}.  The simulations are initialized in a wind-free hydrostatic state using a 1D profile generated by the radiation-convection code {\sc atmo} \citep{tremblin_2015,drummond_2016}.  As the flows found in GCMs can heat the deep atmosphere through transport processes unmodelled in 1D \citep{tremblin_2017}, and as it is prohibitive to run these models until convergence in the deep atmosphere is achieved, we note that there will be a temperature inversion in the final state.  This can be avoided by using a ``hot start" \citep[i.e., using a profile with the tempreature profile artifically shifted to hotter temperatures; ][]{amundsen_2016,tremblin_2017}; however, as our interest is in investigating behaviour seen in \citet{zamyatina_2024}, we opt for the initialization used there.

\subsection{The Parameter Study}

To investigate the impact of the various dissipation mechanisms, we perform simulations of WASP-96b with values from $t_K=1$ to $t_K=12$ (equivalent to $K=0.158$ to $K=0.019$). This range encompasses the default value used in previous studies using the UM \citep[$t_K=1$; see, e.g.,][]{mayne_2019,christie_2021,christie_2022,zamyatina_2023}, the value used in \citet{zamyatina_2024} ($t_K=6$), and the lowest dissipation rate for which the model ran stably for the desired run time of 1000 Earth days ($t_K=12$; for brevity going forward days will refer to Earth days).  While we focus on a sponge depth of $\eta_s=0.75$, as has been used in the previously mentioned papers, we include cases with a raised sponge where $\eta_s=0.9$, with $t_K=1,3$, and $6$.  For the case of $t_K=12$ and $\eta_s=0.9$ the simulation did not run stably.   The impact of the sponge could also be investigated through varying the sponge constant $C$.  Qualitatively, we would expect similar results, as both parameter studies serve to reduce the vertical damping in parts of the atmosphere; however, reducing $C$ explicitly reduces the vertical damping at the outer boundary, unlike altering $\eta_s$, which potentially impacts the effectiveness of the sponge in preventing reflections at the upper boundary. Because of this, we opt to focus on an investigation varying $\eta_s$.

We also note that resolution is expected to influence the dynamics of the jet, both in terms of the implicit dependence within the filtering operator and the numerical limitations of the dynamical core, but also due to the dynamics of smaller scales being resolved with increased resolution impacting the overall results.  While this provides another avenue of investigation and has been explored by \citet{skinner_2021} and \citet{sainsbury-martinez_2019}, we opt to investigate the impact of other parameters instead. 

\begin{table}
\caption{Simulation Parameters}
\label{Tbl:Common}
\begin{tabular}{lcc}
\hline
  & Value & Units\\
\hline
{\em Grid and Domain} \\
Longitude Cells & 144 & \\
Latitude Cells & 90 & \\
Vertical Layers & 66 &\\
Domain Height & $1.03\times 10^7$ & m\\
Hydrodynamic Timestep $\Delta t_\mathrm{dyn}$  & 30 & s \\
\\
{\em Radiative Transfer} \\
Wavelength Bins & 32 & \\
Wavelength Minimum $\lambda_\mathrm{min}$ & 0.2 & $\mathrm{\mu m}$\\
Wavelength Maximum $\lambda_\mathrm{max}$ & 322 & $\mathrm{\mu m}$ \\
Radiative Timestep $\Delta t_\mathrm{rad}$ & 150 & s  \\
\\
{\em Sponge Parameters} \\
Profile & Horizontally Uniform  & \\
Coefficient $C$ & 0.15 & \\
Depth $\eta_s$ & 0.75, 0.90 & \\
\\
{\em Planet}\\
Inner boundary radius & $8.3892\times 10^7$ & m \\
Intrinsic Temperature $T_\mathrm{int}$ & 100 & K \\
Initial Inner Boundary Pressure & 200 & bar \\
Semi-major axis $a$ & $4.53\times 10^{-2}$ & AU \\
Stellar Constant at 1 AU & 1272 & $\mathrm{W\, m^{-2}}$\\
Specific gas constant $R$ & 3164 & $\mathrm{J\,kg^{-1}K^{-1}}$\\
Specific heat capacity $c_\mathrm{P}$ & 11476  & $\mathrm{J\,kg^{-1}K^{-1}}$\\
g at inner boundary & 7.56 & $\mathrm{m\,s^{-2}}$\\
Angular speed $\Omega$ & $2.123\times 10^{-5}$ & $\mathrm{rad\,s^{-1}}$ \\
\hline
\end{tabular}
\end{table}

\section{Results}
\label{Sec:Results}

We present here the results from the suite of simulations investigating the impact of different aspects of the dissipation within the {\sc UM} hot Jupiter model.   Unless otherwise specified, the analysis is done after the pressure, temperature, and velocities have been averaged over the final 100 days of the simulation.  In cases where the vertical coordinate is mapped from a height-based grid to a pressure-based grid, the temporal averaging occurs before the mapping.

\subsection{The Longitudinal Filter}

The motivation for this investigation is a failure in the model to conserve axial angular momentum, resulting in the formation of counter-rotating jets, as seen in \citet{christie_2022} and \citet{zamyatina_2024}. As the primary mechanism for suppressing grid-scale perturbations within hot Jupiter simulations done with the {\sc UM} is the longitudinal filtering of the horizontal velocities and given that that filtering does not necessarily conserve angular momentum, we first examine modifications to the application of the filter and their impact on the jet.

\subsubsection{Reducing the Filter Strength}
\begin{figure*}
	\includegraphics[]{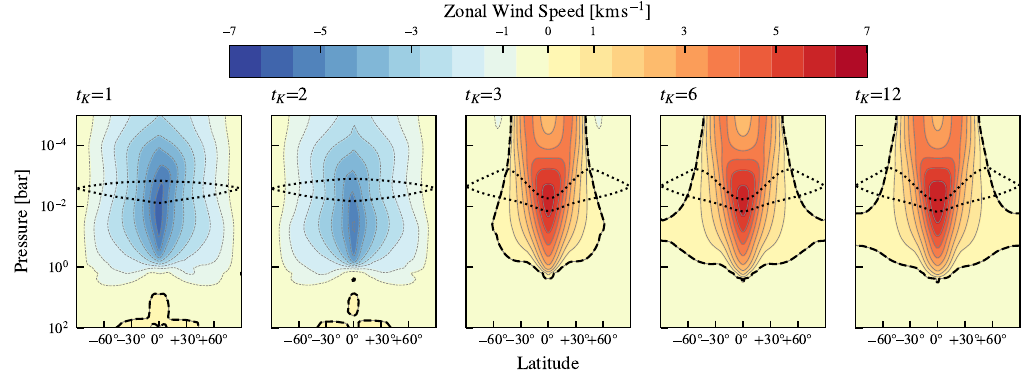}
    \caption{The  mean zonal wind in the $\eta_s=0.75$ models, averaged over the last 100 days.  The dashed line indicates the boundary between super-rotating and counter-rotating flows. The dotted lines indicate the pressure levels at which the pressures partially (lower line) and fully (upper line) intersect the sponge layer.  For the more heavily damped cases ($t_K = 1$ and $2$), a super-rotating jet does not develop.}
    \label{Fig:zonalwind}
\end{figure*}

\begin{figure*}
	\includegraphics[]{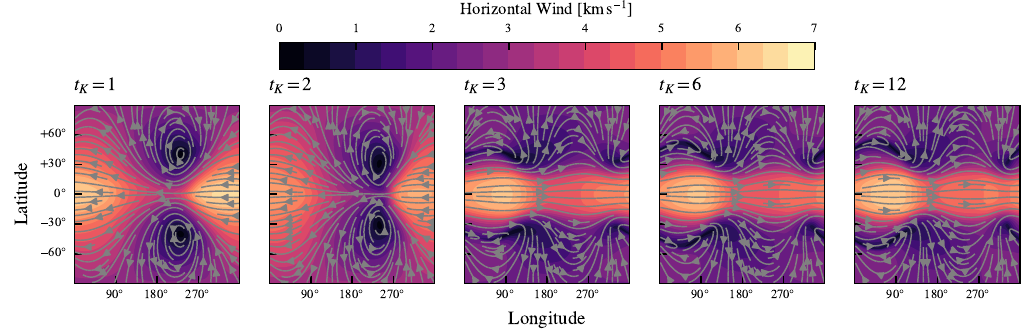}
    \caption{Horizontal wind speed at $1\,\mathrm{mbar}$ averaged between days 900 and 1000 for the $\eta_s=0.75$ cases. The substellar point is located at a longitude of 180\textdegree, the center of each panel. The grey streamlines indicate the direction of the flow. For the more heavily damped cases ($t_K = 1$ and $2$), a super-rotating jet does not develop, instead forming a fast counter-rotating jet. }
    \label{Fig:horizvel}
\end{figure*}

The most direct change that can be made to the filtering is through the e-folding parameter $t_K$.   For the default value $t_K=1$ as well as $t_K=2$, a super-rotating jet forms in the first 100 days; however, this jet quickly erodes, resulting in the majority of the atmosphere counter-rotating, with an equatorial counter-rotating jet (see Figures \ref{Fig:zonalwind} and \ref{Fig:horizvel}).   The flow in these cases forms a pair of dayside vortices on either side of the counter-rotating jet (see Figure \ref{Fig:horizvel}, two leftmost panels), very different from the expected formation of a super-rotating equatorial jet expected from theoretical analysis \citep{showman_2011a,tsai_2014} and seen in GCM simulations of close-in gas giants \citep[e.g.,][]{showman_2008, menou_2012,mayne_2017}.  The dayside flow morphology is reminiscent of the linear flow in \citet{showman_2011a}; however, unlike in that analysis, the nightside is not symmetric.  These two counter-rotating cases show significantly worse conservation of axial angular momentum relative to the remaining cases (see Figures \ref{Fig:aam} and \ref{Fig:dissipation}).  For the $t_K = 1$ case specifically, the simulation loses approximately 3\% of its initial axial angular momentum during the 1000 day run.  A similar breakdown in angular momentum conservation was seen in the $100\times$ solar metallicity, $f_\mathrm{sed}=0.1$ model of GJ 1214b in \citet{christie_2022}.  The rate at which angular momentum is lost during the simulation is not linear in $t_K$ (or $K$), as seen in Figure \ref{Fig:dissipation}, hinting that the loss process might be self reinforcing. These unphysical cases also show discrepant behaviour in their kinetic energy evolution (Figure \ref{Fig:ke}) with the kinetic energy increasing much faster than the cases with super-rotating jets, resulting in the kinetic energy being a factor of $\sim 2$ larger for these cases compared to the super-rotating cases.  Unsurprisingly, these cases show very different thermal structures (see Figures \ref{Fig:pt_equ} and \ref{Fig:pt_midlat}) with less heating deep in the atmosphere and, due to the reversal in the jet, warmer morning terminators and cooler evening terminators.   As these cases exhibit obvious numerical breakdowns and unphysical results, we feel they are easily dismissed and do not dwell on them further.\footnote{The $t_K=1$ and $2$ cases do exhibit a pair of dayside vortices reminiscent of the ``modons'' found in the simulations of \citet{skinner_2023} (cf., their Figure 6c; see also \citealt{skinner_2022a} and \citealt{skinner_2021}); however, as the formation of the vortices in their simulations is robust to changes in dissipation, and given that the GCM used, {\sc pebob}, has been shown to conserve angular momentum in simulations of hot Jupiters \citep{polichtchouk_2014}, the similarities are likely coincidental.}   

In contrast, for the more weakly damped cases ($t_K = 3$, $4$, $6$, and $12$), a zonal jet is seen to form (Figure \ref{Fig:zonalwind}, rightmost three panels) with the expected flow morphology (Figure \ref{Fig:horizvel}, rightmost three panels).  These cases also show improved conservation of the axial angular momentum (Figures \ref{Fig:aam} and \ref{Fig:dissipation}). For comparison, the $t_K=12$ case loses 0.03\% of its initial angular momentum over the 1000 day run, a factor of 100 improvement over the $t_K=1$ case.   The peak zonal velocities approach $6\,\mathrm{km\, s^{-1}}$, with the $t_K=3$ and $4$ cases having the slowest zonal wind at $5.8\,\mathrm{km\, s^{-1}}$.   The counter-rotating regions at low pressures near the poles seen in Figure \ref{Fig:zonalwind} are not counter-rotating jets, but are instead due to the retrograde velocities in the nightside gyres being larger than the prograde velocities on the dayside, as can be seen in Figure \ref{Fig:horizvel}.  Similarly, in the $t_K=6$ and $t_K=12$ cases where the super-rotating region extends to the poles at $P\sim 0.1\,\mathrm{bar}$, this is due to the slower velocities in the gyres and not a jet that extends around the planet.  The centers of the gyres also move towards the poles as suppression of longitudinal variations is reduced.   

\begin{figure*}
	\includegraphics[]{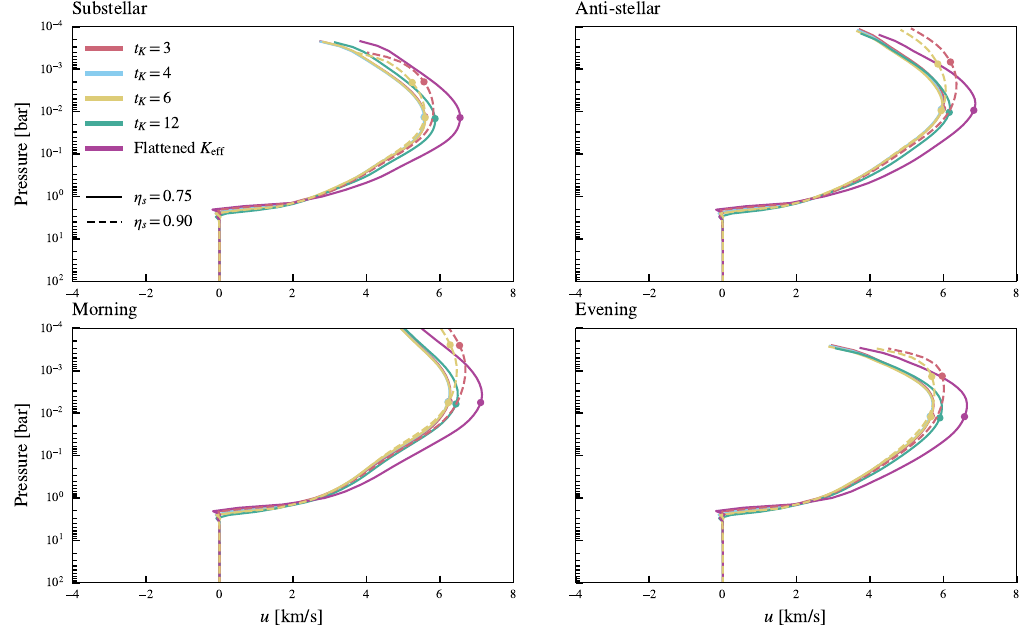}
    \caption{The zonal velocities $u$ at the equator averaged over the final 100 days. The circles indicate where the profiles begin to intersect the sponge.  As the counter-rotating simulations are unphysical, they have been omitted for clarity.}
    \label{Fig:u_equ}
\end{figure*}

\begin{figure*}
	\includegraphics[]{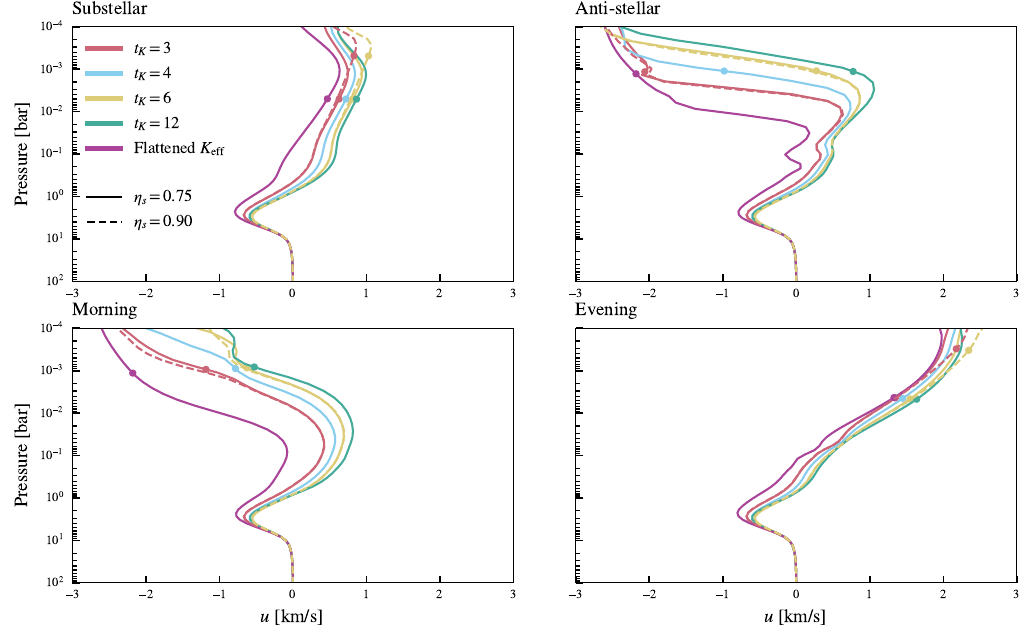}
    \caption{The zonal velocities $u$ at a latitude of $45\degree$  averaged over the final 100 days. The circles indicate where the profiles begin to intersect the sponge.  As the counter-rotating simulations are unphysical, they have been omitted for clarity.}
    \label{Fig:u_midlat}
\end{figure*}

\begin{figure*}
	\includegraphics[]{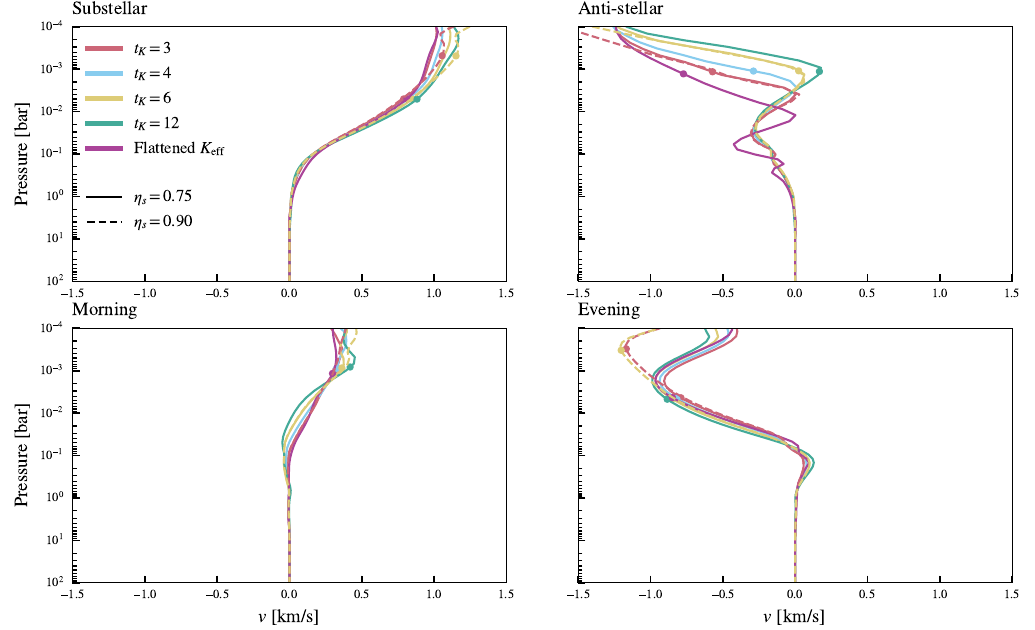}
    \caption{The meridional velocities $v$ at a latitude of $45\degree$  averaged over the final 100 days. The circles indicate where the profiles begin to intersect the sponge.  As the counter-rotating simulations are unphysical, they have been omitted for clarity.}
    \label{Fig:v_midlat}
\end{figure*}

At the equator, the vertical profiles of the local zonal wind (Figure \ref{Fig:u_equ}) do not show significant variation with $t_K$ with the peak of the profiles roughly correlated with the bottom of the sponge. At a latitude of $45\degree$ (Figure \ref{Fig:u_midlat}), however, there is a larger variability in the local zonal wind with $t_K$, typically $\sim 0.5\,\mathrm{km\,s^{-1}}$, with the largest dependency on $t_K$ seen in the anti-stellar and morning profiles.  At the mid-latitudes, unlike at the equator, the peak of the profiles do not correspond to the bottom of the sponge and show a more complex, longitudinally-dependent vertical behaviour\footnote{While the plots show only profiles at a latitude of $45\degree$, the results are qualitatively similar at other latitudes outside of the jet.}. The meridional velocities (Figure \ref{Fig:v_midlat}) show some variability at a latitude of $45\degree$, especially in the anti-stellar profile where, for example, at a pressure of 1 mbar the $t_K=3$ and $t_K=12$ velocities differ by $0.5\,\mathrm{km\,s^{-1}}$, in part due to changes in the direction -- northward to southward -- occurring at different pressures.  The local dependence of the meridional velocity on $t_K$, however, varies greatly with the choice of latitude.  A similar examination of the meridional velocities as in Figure \ref{Fig:v_midlat}, except at a latitude of $60\degree$ (not shown), finds improved agreement in the anti-stellar profile but larger variability with $t_K$ in the profile along the morning terminator.  In general, greater variability with $t_K$ is seen in the horizontal winds outside of the equatorial jet compared to variability of the zonal velocities within the jet. 

\begin{figure*}
	\includegraphics[]{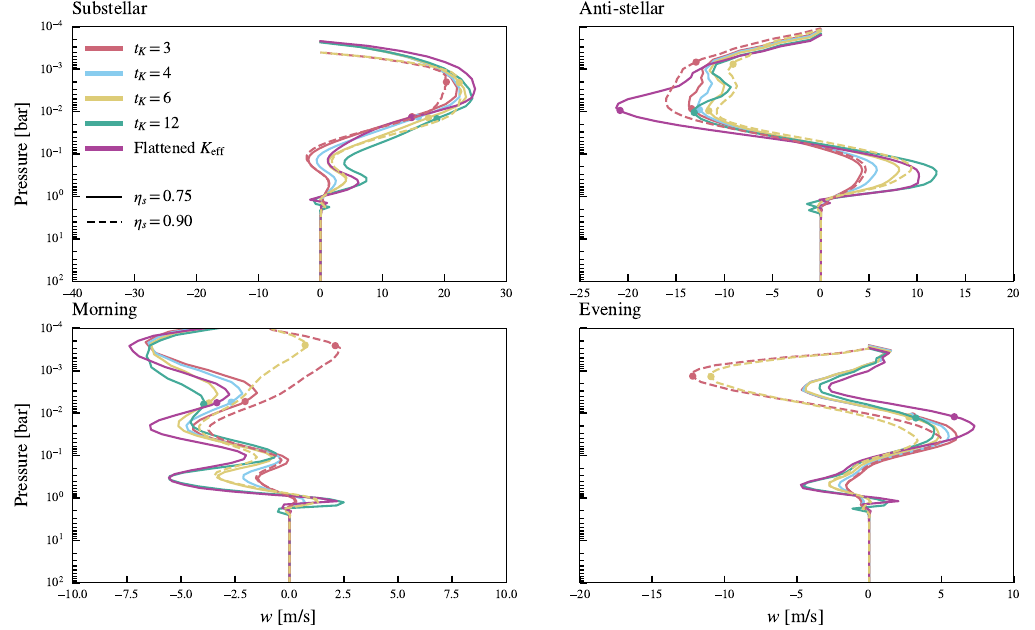}
    \caption{The vertical velocities $w$ at the equator averaged over the final 100 days. The circles indicate where the profiles begin to intersect the sponge.  As the counter-rotating simulations are unphysical, they have been omitted for clarity.}
    \label{Fig:vert_vel}
\end{figure*}

The equatorial vertical velocities for these cases are shown in Figure \ref{Fig:vert_vel}.    The velocities within the equatorial jet do not show indication of agreement between $0.1$ and $1\,\mathrm{bar}$ with the vertical velocities in the $t_K=12$ case being up to $\sim 2\times$ larger than the vertical velocities in the $t_k=6$ case. This region is just above the temperature peak seen in Figure \ref{Fig:pt_equ} and shows the impact of horizontal filtering on the vertical transport of potential temperature \citep{tremblin_2017,sainsbury-martinez_2019}.   Were the initial pressure-temperature profile used within the simulations to be closer to the final, converged state,  or if a ``hot start'' were used \citep{amundsen_2016,tremblin_2017}, it is unclear as to whether these differences in this region would have been seen.  The vertical velocities at the equator also show indications of not being converged in the morning profiles between $10^{-2}$ and $10^{-3}\,\mathrm{bar}$ with differences in this region between the $t_K=6$ and $t_K=12$ cases being up to $\sim 20\%$.

The pressure-temperature profiles at the equator and at a latitude of $45\degree$ are shown in Figures \ref{Fig:pt_equ} and \ref{Fig:pt_midlat}, respectively.   For the cases where $t_K = 3$, $4$, $6$, and $12$, the temperature profiles show good agreement except at pressures $\lesssim 10^{-2}\,\mathrm{bar}$ on the nightside and morning terminator.   This region corresponds to the nightside gyres which, due to the lack of external radiative forcing and the longer dynamical timescales, are impacted more by the small velocity perturbation allowed by the weaker filtering.  As discussed above, the temperatures at $\sim 1\,\mathrm{bar}$ also increase with reduced filtering as the transport of potential temperature into the deeper atmosphere is enhanced.  

While we focus on $t_K$ as the natural parametrisation for the dissipation from a numerical point of view, this approach is resolution dependent (see Eqn. \ref{Eqn:Keff}) with $K_\mathrm{eff}$ being the dynamically relevant quantity.  In Appendix \ref{Apdx:Res} we test that a sufficient increase the horizontal resolution, corresponding to a reduction of $K_\mathrm{eff}$, does result in the formation of a super-rotating jet for case of $t_K=1$.

\begin{figure*}
	\includegraphics[]{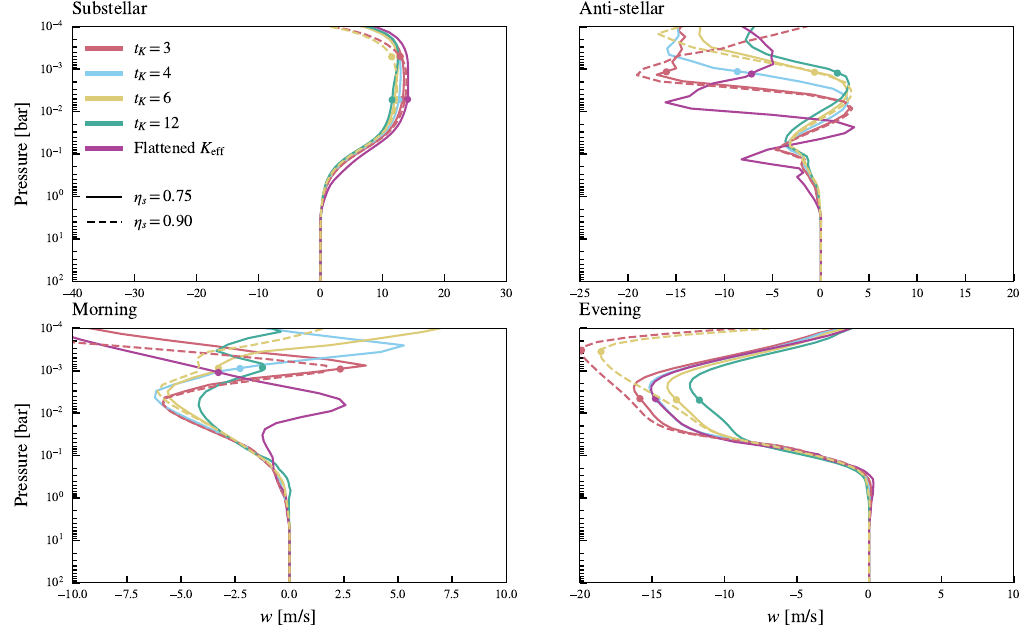}
    \caption{The vertical velocities $w$ at a latitude of $45\degree$ averaged over the final 100 days. The circles indicate where the profiles begin to intersect the sponge. As the counter-rotating simulations are unphysical, they have been omitted for clarity.}
    \label{Fig:vert_vel_midlat}
\end{figure*}

\begin{figure*}
	\includegraphics[]{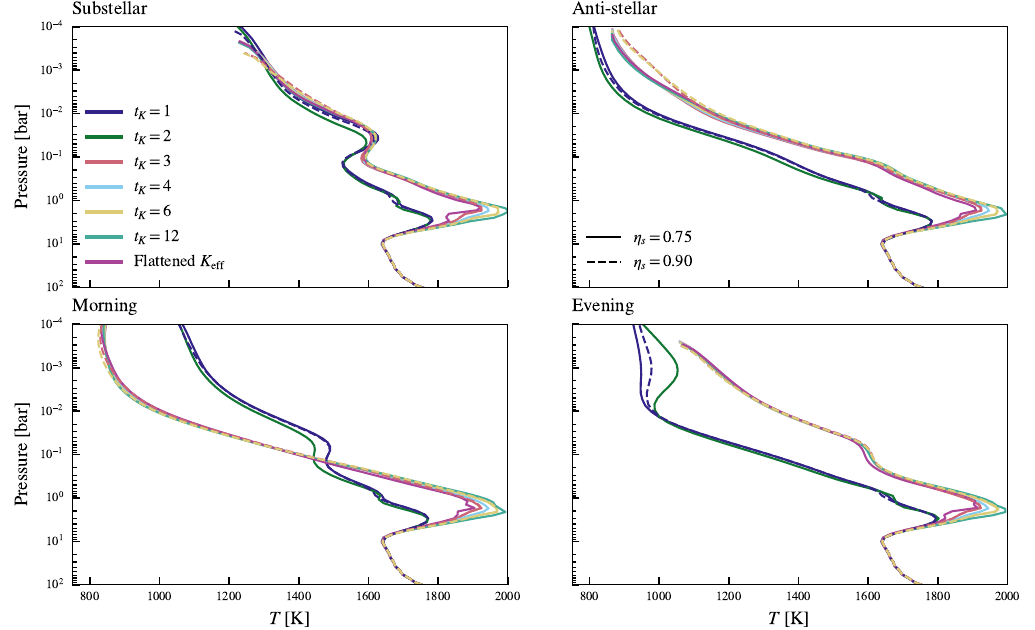}
    \caption{Temperature profiles at the equator, averaged over the last 100 days.  Colors indicate the value of $t_K$ (see legend in the upper left panel) while linestyles indicate the sponge parameter (solid lines correspond to $\eta_s=0.75$ while dashed lines correspond to $\eta_s=0.9$).}
    \label{Fig:pt_equ}
\end{figure*}

\begin{figure*}
	\includegraphics[]{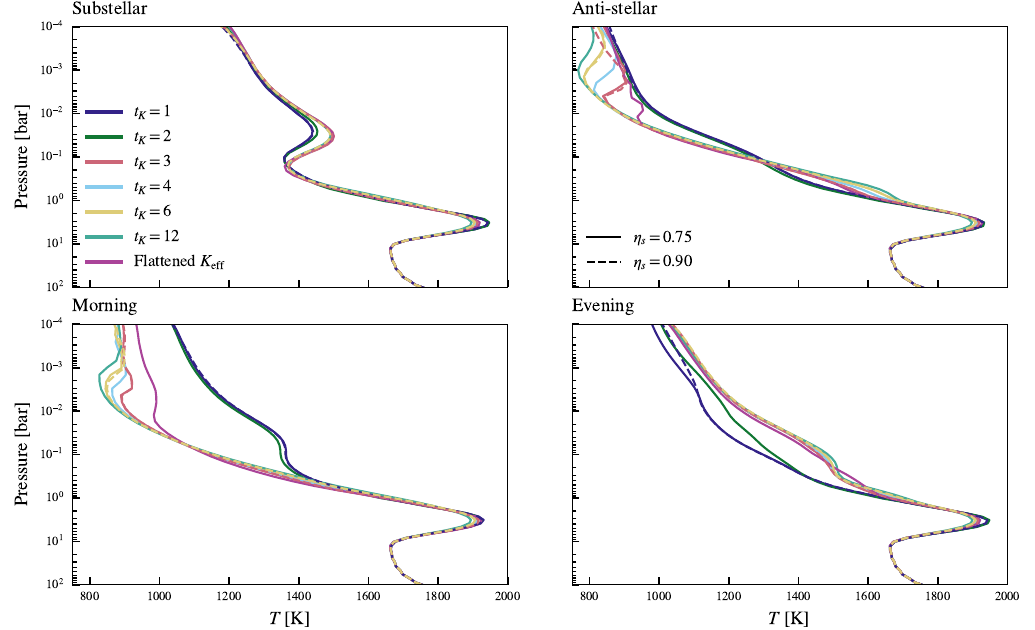}
    \caption{Temperature profiles at a latitude of $45\degree$, averaged over the last 100 days.  As with Figure \ref{Fig:pt_equ}, the colors indicate the value of $t_K$ (see legend in the upper left panel) while linestyles indicate the sponge parameter (solid lines correspond to $\eta_s=0.75$ while dashed lines correspond to $\eta_s=0.9$.}
    \label{Fig:pt_midlat}
\end{figure*}

\begin{figure}
	\includegraphics[]{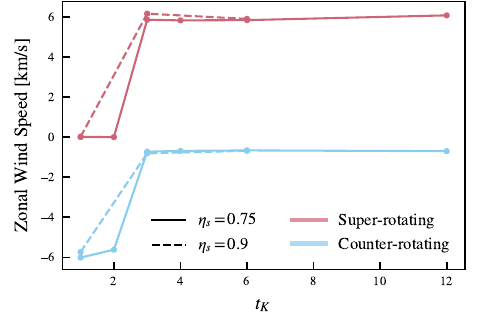}
    \caption{The peak mean super-rotating (red) and counter-rotating (blue) zonal winds for each value of $t_K$ and $\eta_s$. For comparison, the flattened $K_\mathrm{eff}$ simulation had a peak zonal velocity of $6.79\,\mathrm{km\,s^{-1}}$.}
    \label{Fig:zwlp}
\end{figure}

\begin{figure}
	\includegraphics[]{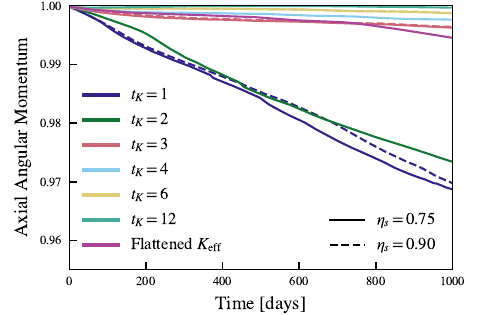}
    \caption{The axial angular momentum evolution for each case, normalized to the initial axial angular momentum.}
    \label{Fig:aam}
\end{figure}

\begin{figure}
	\includegraphics[]{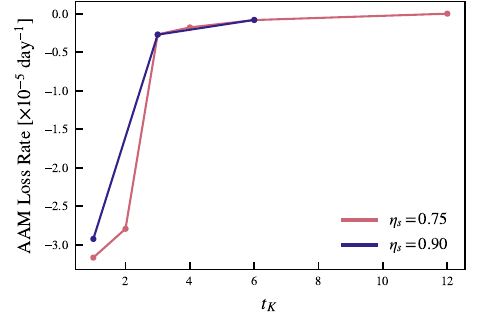}
    \caption{Rate of dissipation of the normalized axial angular momentum based on a linear fit to the axial angular momentum evolution in Figure \ref{Fig:aam}.}
    \label{Fig:dissipation}
\end{figure}

\begin{figure}
	\includegraphics[]{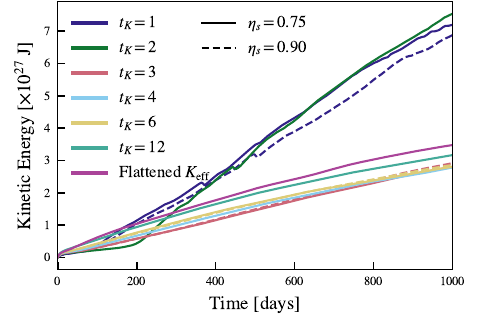}
    \caption{The evolution of the kinetic energy for each simulation.  The simulations which resulted in counter-rotating jets show a significantly faster growth in their total kinetic energy.}
    \label{Fig:ke}
\end{figure}

\subsubsection{Flattening the diffusion profile}
\label{Sec:Flat}
As, by design, the longitudinal filtering acts on the grid without consideration of the metric, the effective diffusion coefficient varies with latitude (see Equation \ref{Eqn:Keff}) which has the potential for impacting the equatorial jet.   To investigate this possibility, we run a simulation with $\eta_s=0.75$ and change the filter coefficient K to partially account for the implicit latitudinal gradient:
\begin{equation}
K = \frac{1}{4}\left(1-e^{-\frac{1}{3}}\right)\min\left(1, \frac{\cos^2\phi_0}{\cos^2\phi} \right)\,\, ,
\end{equation}

\noindent where $\phi_0 = 57.92\degree$. This choice corresponds to $t_K=3$ in the polar caps $|\phi| >\phi_0$ while holding the effective diffusion coefficient ($K_\mathrm{eff}$) constant in the equatorial region $-\phi_0 < \phi < \phi_0$, resulting in $t_K=12$ at $\phi=0\degree$.  While it may be possible to run the simulation with a larger region of constant $K_\mathrm{eff}$, this choice restricts the filtering to values that have been investigated and shown to form a super-rotating jet, facilitating the interpretation of results as it is less likely that any differences will be due to extreme choices in the local filtering.

The zonal velocities are shown in Figure \ref{Fig:zonalwind_eta9} (rightmost panel).  The peak zonal velocity is 6.79 $\mathrm{km\,s^{-1}}$, faster than for the $t_K=3$ ($5.87\, \mathrm{km\,s^{-1}}$) and $t_K=12$ ($ 6.09\,\mathrm{km\,s^{-1}}$) filtering profiles which bracket the flattened $K_\mathrm{eff}$ profile;  moreover, the vertical profiles of the zonal velocity (Figure \ref{Fig:u_equ}) show the zonal velocities to be consistently larger than in any of the other $t_K$ values examined with $\eta_s=0.75$. This indicates that while there is an increase in zonal velocity with the reduced filter strength, the profile itself impacts the velocity to a greater degree than the strength, at least within the range of values required for numerical stability.  

The temperature structure is impacted in the same regions as seen in the $t_K$ parameter study investigated in the previous section:  the equatorial temperatures are relatively insensitive to the filtering, except at $P \sim 1\,\mathrm{bar}$ where the downward transport of entropy results in deep atmosphere heating and the previously discussed temperature inversion.    At the mid-latitudes (Figure \ref{Fig:pt_midlat}), the temperatures are hotter at $P \lesssim 10^{-2}\,\mathrm{bar}$ around the nightside and morning terminator -- the region of the nightside gyres -- compared to the $t_K=3$ case, where it was seen in the previous section that weaker filtering resulted in cooler mid-latitude temperatures in the same region.  This reinforces the idea that relative to the jet, the temperatures in the gyres are more sensitive to the local, smaller-scale transport processes.

\subsection{The Sponge}
The fiducial suite of simulations were performed with the sponge parameter of $\eta_s=0.75$, i.e., the sponge covers the upper quarter of the computational domain.    As the sponge strength is parametrised based on height (see Equation \ref{Eqn:Sponge}), the sponge strength is not constant along isobaric surfaces, and for some pressures the isobaric surface only partially intersects the sponge.   This can be seen in Figure \ref{Fig:zonalwind} where the lower dotted lines indicate where the surface begins to intersect the sponge and the upper dotted line indicates the lowest point where the isobaric surface is entirely within the sponge.  Due to the sponge extending into the jet, reaching, at least partially, pressures of $10^{-2}$ bar, we opt to investigate the impact of moving the sponge upward to $\eta_s=0.9$. \footnote{For this value of $\eta_s$ and vertical grid, the sponge only occupies six grid zones, making a further reduction unlikely.  Another option would be to increase $z_\mathrm{top}$ while adding additional grid zones to move the sponge upward; however, that is not investigated here.}

\begin{figure*}
	\includegraphics[]{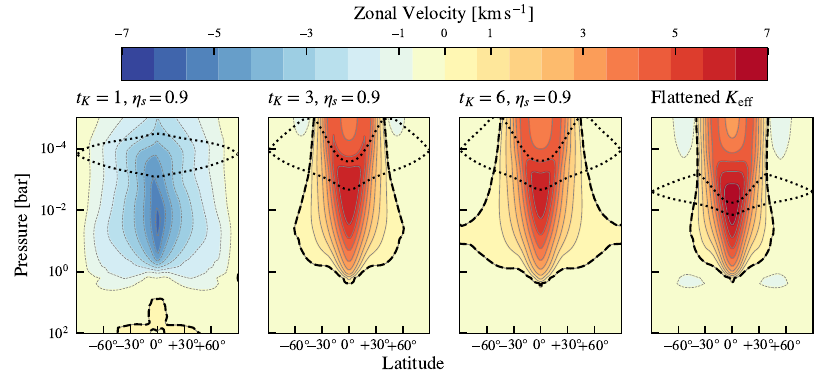}
    \caption{The mean zonal wind in the $\eta_s=0.9$ simulations (left three panels), and the simulation with a flattened $K_\mathrm{eff}$ and $\eta_s=0.75$ (rightmost panel) , averaged over the last 100 days.  The dashed line indicates the boundary between super-rotating and counter-rotating flows. The dotted lines indicate the pressure levels at which the pressures partially (lower line) and fully (upper line) intersect the sponge layer.  For the more heavily damped cases ($t_K = 1$ and $2$), a super-rotating jet does not develop.}
    \label{Fig:zonalwind_eta9}
\end{figure*}

\begin{figure*}
	\includegraphics[]{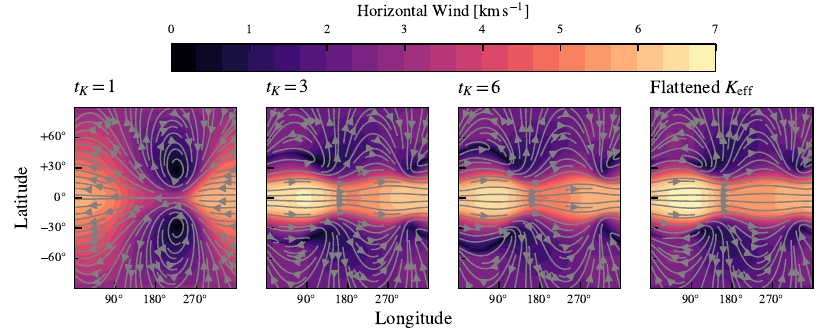}
    \caption{Horizontal wind speed at $1\,\mathrm{mbar}$ averaged between days 900 and 1000 for the $\eta_s=0.9$ cases (three leftmost panels) and the flattened $K_\mathrm{eff}$ profile case with $\eta_s=0.75$ (rightmost panel). The substellar point is located at a longitude of 180\textdegree, the center of each panel. The grey streamlines indicate the direction of the flow. }
    \label{Fig:horizvel9}
\end{figure*}

The zonal velocities for the simulations with $\eta_s=0.9$ are shown in Figure \ref{Fig:zonalwind_eta9} and the horizontal velocities at 1 mbar are shown in Figure \ref{Fig:horizvel9}.   The case of $t_K=1$ still forms a counter-rotating jet with a breakdown in the conservation of axial angular momentum (Figure \ref{Fig:aam}) and an excess of kinetic energy relative to the less dissipative cases (Figure \ref{Fig:ke}). The rate of axial angular momentum loss and the growth rate of the kinetic energy show larger differences between the two sponge cases for $t_K=1$ compared to the super-rotating cases where we see a greater degree of agreement between the sponge cases; however, the resulting flow morphology and temperature structure are largely the same for both $t_K=1$ cases.  The cases which form a super-rotating jet ($t_K=3$ and $t_K=6$), a similar level of dissipation (Figure \ref{Fig:dissipation}) is seen compared to the deep sponge ($\eta_s=0.75$) cases, indicating that the sponge is not indirectly influencing the rate of axial angular momentum loss to a significant degree.   The $t_K=3$ case shows a faster peak zonal velocity than the $t_K=6$ case ($6.18\,\mathrm{km\,s^{-1}}$ compared to $5.92\,\mathrm{km\,s^{-1}}$; see Figure \ref{Fig:zwlp}), while in the deep sponge case the $t_K=3$ and $t_K=6$ cases have essentially the same peak zonal windspeed ($5.87\,\mathrm{km\,s^{-1}}$ and $5.85\,\mathrm{km\,s^{-1}}$, respectively). The zonal velocities also begin to decrease with pressure higher in the atmosphere compared to the deep sponge case, indicating that the sponge serves to suppress the jet.  Although not investigated here, the zonal wind speeds are likely also impacted by the upper boundary as the the maximum equatorial pressures at the upper boundary are between $0.3\,\mathrm{mbar}$ and $0.45\,\mathrm{mbar}$, depending on the  simulation.  Equatorial zonal wind speeds at pressures lower than this are certainly altered by the upper boundary; however, the exact impact at larger pressures is unclear. While varying the location of the upper boundary is beyond the scope of this parameter study investigated here, we remain cognizant of its potential impact on the results.  

The temperature differences between the two sponge cases are small, with the largest differences occurring around the anti-stellar point at the equator where the temperatures differ by $\lesssim 50\,\mathrm{K}$ for pressures $\lesssim 8\times 10^{-2}\,\mathrm{bar}$. This is expected since as the nightside is cooler and thus has smaller scale heights, the sponge does not reach the high pressures it does on the dayside due to the sponge layer being parametrised based on height, reducing its impact on the flow.   The sponge can be seen to have a larger influence, perhaps unsurprisingly, on the vertical velocity $w$. At the terminators along the equator (Figure \ref{Fig:vert_vel}) , the velocities at pressures $\lesssim 10^{-2}\,\mathrm{bar}$ can differ by up to $7\,\mathrm{m\,s^{-1}}$, comparable to the peak vertical velocities seen in those regions.  Moreover,  sharp transitions to the sponge and the damping of the vertical velocities to zero can be seen at low pressures. This is especially apparent at mid-latitudes along the evening terminator (see Figure \ref{Fig:vert_vel_midlat}, lower right panel).   While the temperatures are not significantly altered by the sponge, physical processes not considered here which depend on the local transport properties (e.g., disequilibrium chemistry, aerosols) may have their results altered by its presence.  Similarly, post-processed models that make use of the local vertical velocities extracted from GCMs \citep[e.g.,][]{tsai_2023} may also include impacts from the nature of the sponge applied in the original simulations.  

\subsection{Other Impacts of Damping}

While the cases of $t_K=1$ and $t_K=2$ can safely be discounted as unphysical, the differing levels of damping within the subset of simulations that do form a super-rotating jet allows for the impact of the damping to be better understood.  In this subsection we look at a few other aspects of the simulations affected by dissipation.

\subsubsection{Spin Up}

\begin{figure}
	\includegraphics[]{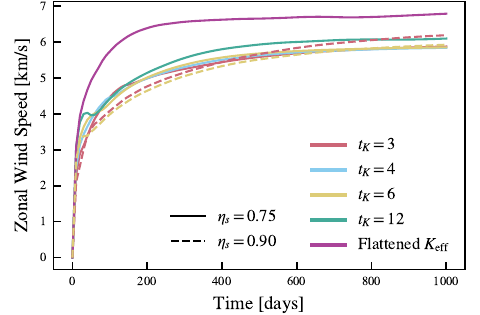}
    \caption{The evolution of the peak zonal wind speed for each case that resulted in a super-rotating jet. }
    \label{Fig:zwts}
\end{figure}

Since the choice is made to run the simulation for a fixed time period as running to convergence might not be feasible, the spin-up period cannot be discounted in determining the state at the end of the simulation.  Figure \ref{Fig:zwts} shows the evolution of the instantaneous peak zonal wind, sampled every 10 days, over the length of each simulation.   Within the first 200 days, the zonal winds oscillate in peak amplitude, with the less damped simulations having larger oscillations.   Within this early phase, up until day 400, the deep sponge simulations ($\eta_s=0.75$) spin up faster; however, at day 1000 the shallow sponge simulations ($\eta_s=0.9$) have faster zonal winds than those simulations with the same $t_K$.   There is not, unfortunately, an obvious trend with $t_K$ with the $t_K=12$ simulation having the fastest zonal wind for the deep sponge cases and $t_K=3$ having the fastest zonal wind for the shallow sponge cases.  The simulation with a flattened $K_\mathrm{eff}$ stands out from the rest of the simulations as spinning up faster than any model with a fixed $t_K$ and having the fastest jet after 1000 days.    In all cases, the jets continue to accelerate, with rates on the order of a few $\times 10^{-4}\,\mathrm{km\, s^{-1}\, day^{-1}}$, based on an estimate of the rate of change over the final 100 days of each simulation.   As with the jet speeds, there is not an obvious trend in the jet acceleration rate with $t_K$; however, the simulations with a shallow sponge do have higher zonal wind accelerations in the last 100 days than those with a deep sponge. For the shallow sponge, the $t_K=3$ and $t_K=6$ cases are accelerating at $5.7 \times 10^{-4}\,\mathrm{km\, s^{-1}\, day^{-1}}$ and $4.6 \times 10^{-4}\,\mathrm{km\, s^{-1}\, day^{-1}}$ while for the deep sponge the $t_K=3$ case accelerates the fastest at the end at $2.9 \times 10^{-4}\,\mathrm{km\, s^{-1}\, day^{-1}}$.  For comparison, the simulation with the flattened $K_\mathrm{eff}$ had an acceleration of $4.8 \times 10^{-4}\,\mathrm{km\, s^{-1}\, day^{-1}}$.   This indicates that while the shallow sponge may result in a faster continued acceleration of the jet, the integration times required to see a significant increase in the jet speed based on a linear extrapolation are prohibitive. 

\begin{figure*}
	\includegraphics[]{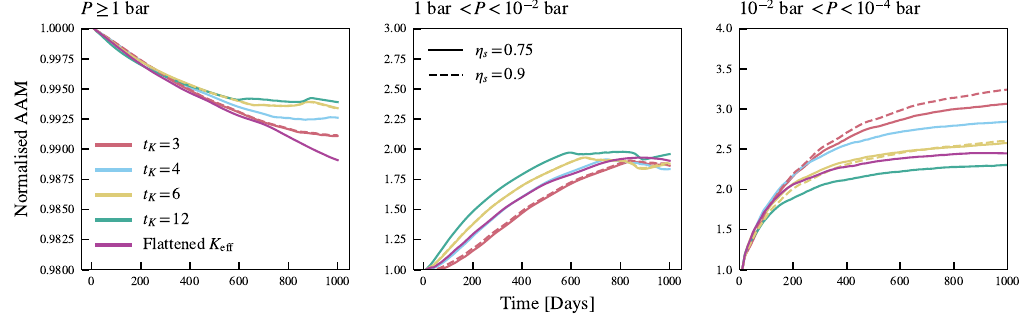}
    \caption{The evolution of the axial angular momentum within three regions of the atmosphere.  In each case, the axial angular momentum within the layer is normalised to the value at the beginning of the simulation.  }
    \label{Fig:aam_layers}
\end{figure*}

\begin{figure*}
	\includegraphics[]{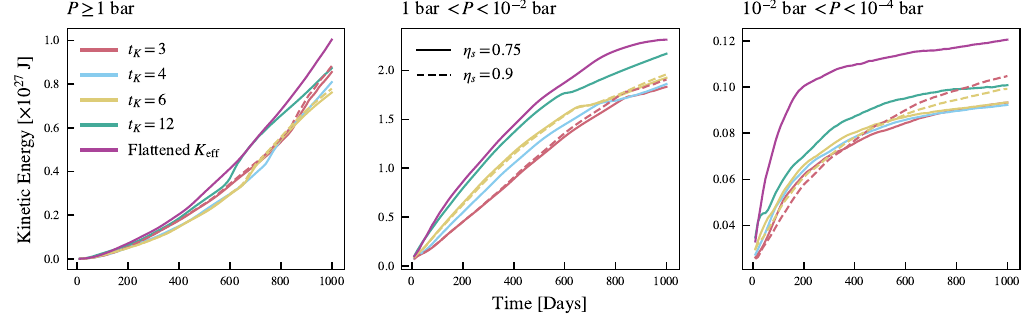}
    \caption{The evolution of kinetic energy within three regions of the atmosphere.  Unlike the axial angular momentum in Figure \ref{Fig:aam_layers}, no normalisation is applied. }
    \label{Fig:ke_layers}
\end{figure*}

We additionally look at the evolution of the axial angular momentum at fixed pressure intervals over the duration of the simulation (see Figure \ref{Fig:aam_layers}).   The lowest pressure interval is $10^{-2}\,\mathrm{bar} < P < 10^{-4}\,\mathrm{bar}$.  At even lower pressures, isobaric surfaces in some of the simulations intersect the upper boundary, complicating the interpretation of the results.  The $t_K=1$ and $t_K=2$ cases are also excluded as loss of angular momentum makes comparison difficult\footnote{We also note that in the upper atmosphere for these cases that the sign of the axial angular momentum changes as the relative counter-rotation exceeds the rotational velocity of the planet.}. As is shown in the Figure \ref{Fig:aam_layers} (left panel), the deep ($P > 1\,\mathrm{bar}$) atmosphere loses axial angular momentum as the jet above forms, resulting in the deep atmosphere to slowly counter-rotate (see Figures \ref{Fig:zonalwind} and \ref{Fig:zonalwind_eta9} for the zonal winds).  While some of this can be attributed to the global loss of angular momentum, the trend is seen in all the simulations, including the $t_K=6$ and $t_K=12$ cases which have improved conservation properties, indicating that this effect is predominantly due to the transport of angular momentum to lower pressures.  These latter cases exhibit a leveling off of the axial angular momentum by day 600 both in the deep atmosphere and in the intermediate atmosphere ($1\,\mathrm{bar} < P < 10^{-2}\,\mathrm{bar}$). While this could be interpreted as the upper atmosphere approaching a dynamically steady state, the atmospheric kinetic energy (Figure \ref{Fig:ke_layers}), although showing some slowing, continues to evolve.  We thus conclude that even with the shorter evolutionary timescales of the upper atmosphere relative to the deep atmosphere, the upper atmosphere in the simulation has not fully converged. 

\subsubsection{Vertical mixing}

Extracting an effective vertical mixing parameter $K_{zz}$ from GCM simulations has become an important input to one- and two-dimensional models of higher microphysical complexity \cite[e.g.,][]{powell_2019,tsai_2024b}.   While tracer-based methods \citep[e.g.,][]{parmentier_2013} may provide more precise estimates of the global mixing, the analysis of \citet{komacek_2019} provides a straightforward estimate of $K_{zz}$,
\begin{equation}
K_{zz} \sim \frac{\overline{w}^2}{\tau_\mathrm{adv}^{-1} + \tau_\mathrm{chem}^{-1}} \sim \overline{w}^2\tau_\mathrm{adv}\, ,
\label{Eqn:Kzz}
\end{equation}

\noindent where $\overline{u}$ and $\overline{w}$ are the root-mean-squared zonal and vertical velocities, computed on isobaric surfaces, and the advection timescale is estimated as $\tau_\mathrm{adv} = R_\mathrm{p}/\overline{u}$. The radius $R_\mathrm{p}$ is taken to be the inner boundary radius (see Table \ref{Tbl:Common}). For the purposes of this estimate, we ignore the chemical timescale $\tau_\mathrm{chem}$ and only consider extremely long-lived tracers in the atmosphere.  We also note that this reduces to the traditional estimate of vertical mixing, $K_{zz} \sim \overline{w} H$, where $H$ is the scale height, when $\overline{u}/R_\mathrm{p}\sim \overline{w}/H$.  This has been shown to overestimate the global mixing \citep{parmentier_2013}; however it has been used, with an overall scaling applied, when an estimate of the local mixing rate is needed, as opposed to a global value \citep{tsai_2023}.   

\begin{figure}
	\includegraphics[]{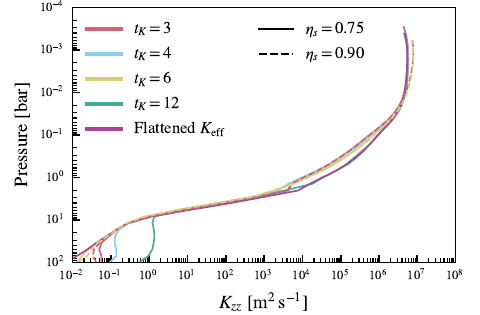}
    \caption{$K_{zz}$ estimated for each simulation using Eqn. \ref{Eqn:Kzz}, computed using the average velocities and pressures over the final 100 days of the simulations.  The simulations which resulted in counter-rotating jets have been omitted as they represent unphysical results.}
    \label{Fig:kzz}
\end{figure}

Figure \ref{Fig:kzz} shows the estimates for $K_{zz}$ based on Equation \ref{Eqn:Kzz}.   We find that reduced vertical damping by the sponge leads to a 60\% increase in $K_{zz}$ above $10^{-2}\, \mathrm{bar}$, which, given that the parameter space of mixing rates can span many orders of magnitude, is a relatively modest increase.  At pressures greater than $10$ bar, larger differences are present, with the simulations with increased damping showing order of magnitude decreases in vertical mixing. This region, however, is not converged in the simulation, as shown in Figure \ref{Fig:pt_equ} where the region below 10 bar has yet to be heated by the downward advection of entropy \citep{tremblin_2017}.  We further note that the $K_{zz}$ profiles computed from instantaneous  outputs as opposed to a time-average show even larger variations in the mixing at pressures larger than 10 bar, while at pressures less than 10 bar the $K_{zz}$ estimates show little variation relative to their time-averaged values.

\section{Conclusions}
\label{Sec:Conclusions}

In this paper, we demonstrate that sufficiently strong longitudinal filtering of the horizontal velocity can lead to a suppression of a zonal equatorial jet and a breakdown in the conservation of axial angular momentum; however, for the case of WASP-96b investigated here, we find that once the filter is weak enough for a super-rotating jet to form, the speed of the jet and the thermal structure are relatively insensitive to the filter coefficient $K$.  An exception to this is within the lower pressure ($\lesssim 10^{-2}\,\mathrm{bar}$) nightside gyres where the temperatures can vary with $K$ up to $\sim 100\,\mathrm{K}$.  A larger influence on the jet is seen from the latitudinal dependence of the filtering.  As a latitude-longitude grid is used within the model, and since the filtering acts on the grid directly without involving the metric, filtering with a constant value of $K$ results in stronger diffusion at latitudes near the equator.   By altering $K$ to partially account for this, resulting in a constant effective diffusion parameter $K_\mathrm{eff}$ near the equator, we show that the latitudinal profile of $K$ can impact the zonal jet to a greater degree than the choice of $t_K$, at least within the range of values that maintain numerical stability.  We also find that the choice of dissipation can impact the vertical velocities, even in regions where only minor temperatures differences are seen.   This may affect chemical transport processes, although that was not investigated in this paper.   Given the impact of such dissipation on the simulations, observations are required to constrain the magnitude of the combined numerical and physical dissipative mechanisms. Additionally, as there are several physical processes likely to impart dissipation which are not explicitly included in GCM simulations (e.g., magnetic drag, gravity waves, shocks), although instructive, higher spatial and temporal resolution simulations do not provide an ultimate `ground truth' nor remove the requirement for observational constraint to determine the dynamical structures of gas giant atmospheres. 

\section*{Acknowledgements}
D.A.C. is supported by funding from the Max Planck Society. This research was also supported by a UK Research and Innovation (UKRI) Future Leaders Fellowship MR/T040866/1, and partly supported by the Leverhulme Trust through a research project grant RPG-2020-82 alongside a Science and Technology Facilities Council (STFC) Consolidated Grant ST/R000395/1.  H. B. was supported by the Engineering and Physical Sciences Research Council (EPSRC) Vacation Internship Scheme 2022.

The simulations in this work were performed using the DiRAC Data Intensive service at Leicester, operated by the University of Leicester IT Services, which forms part of the STFC DiRAC HPC Facility (www.dirac.ac.uk). The equipment was funded by BEIS capital funding via STFC capital grants ST/K000373/1 and ST/R002363/1 and STFC DiRAC Operations grant ST/R001014/1. DiRAC is part of the National e-Infrastructure.  As supercomputing is an energy-intensive endeavor, we note that the production runs used in this paper required  241496 CPU-hours, resulting in 500 kg of \ce{CO2} being emitted, based on estimates for UK electricity generation\footnote{https://www.gov.uk/government/publications/greenhouse-gas-reporting-conversion-factors-2023}.  This excludes emissions due to testing and analysis and is certainly an underestimation of the carbon impact of this research.  

The analysis of the simulation data and the production of plots for this paper made use of the following Python packages: {\sc aeolus} \citep{sergeev_2022}, {\sc iris} \citep{hattersley_2023}, {\sc matplotlib} \citep{hunter_2007}, and {\sc numpy} \citep{harris_2020}.  

The authors would also like to thank the anonymous referee for their comments which greatly improved the quality of the paper.

\section*{Data Availability}

The simulation data used in this paper are available from the Zenodo online repository at \href{https://doi.org/10.5281/zenodo.11451199}{doi.org/10.5281/zenodo.11451199}.  For the purpose of open access, the authors have applied a Creative Commons Attribution (CC BY) licence to any Author Accepted Manuscript version arising.



\bibliographystyle{mnras}
\bibliography{polarfilter} 

\begin{thebibliography}{}
\makeatletter
\relax
\def\mn@urlcharsother{\let\do\@makeother \do\$\do\&\do\#\do\^\do\_\do\%\do\~}
\def\mn@doi{\begingroup\mn@urlcharsother \@ifnextchar [ {\mn@doi@}
  {\mn@doi@[]}}
\def\mn@doi@[#1]#2{\def\@tempa{#1}\ifx\@tempa\@empty \href
  {http://dx.doi.org/#2} {doi:#2}\else \href {http://dx.doi.org/#2} {#1}\fi
  \endgroup}
\def\mn@eprint#1#2{\mn@eprint@#1:#2::\@nil}
\def\mn@eprint@arXiv#1{\href {http://arxiv.org/abs/#1} {{\tt arXiv:#1}}}
\def\mn@eprint@dblp#1{\href {http://dblp.uni-trier.de/rec/bibtex/#1.xml}
  {dblp:#1}}
\def\mn@eprint@#1:#2:#3:#4\@nil{\def\@tempa {#1}\def\@tempb {#2}\def\@tempc
  {#3}\ifx \@tempc \@empty \let \@tempc \@tempb \let \@tempb \@tempa \fi \ifx
  \@tempb \@empty \def\@tempb {arXiv}\fi \@ifundefined
  {mn@eprint@\@tempb}{\@tempb:\@tempc}{\expandafter \expandafter \csname
  mn@eprint@\@tempb\endcsname \expandafter{\@tempc}}}

\bibitem[\protect\citeauthoryear{Amundsen, Baraffe, Tremblin, Manners, Hayek,
  Mayne  \& Acreman}{Amundsen et~al.}{2014}]{amundsen_2014a}
Amundsen D.~S.,  Baraffe I.,  Tremblin P.,  Manners J.,  Hayek W.,  Mayne
  N.~J.,   Acreman D.~M.,  2014, \mn@doi [Astronomy \& Astrophysics]
  {10.1051/0004-6361/201323169}, 564, A59

\bibitem[\protect\citeauthoryear{Amundsen et~al.,}{Amundsen
  et~al.}{2016}]{amundsen_2016}
Amundsen D.~S.,  et~al., 2016, \mn@doi [Astronomy \& Astrophysics]
  {10.1051/0004-6361/201629183}, 595, A36

\bibitem[\protect\citeauthoryear{Amundsen, Tremblin, Manners, Baraffe  \&
  Mayne}{Amundsen et~al.}{2017}]{amundsen_2017}
Amundsen D.~S.,  Tremblin P.,  Manners J.,  Baraffe I.,   Mayne N.~J.,  2017,
  \mn@doi [Astronomy \& Astrophysics] {10.1051/0004-6361/201629322}, 598, A97

\bibitem[\protect\citeauthoryear{Beltz, Rauscher, Roman  \& Guilliat}{Beltz
  et~al.}{2022}]{beltz_2022}
Beltz H.,  Rauscher E.,  Roman M.~T.,   Guilliat A.,  2022, \mn@doi [The
  Astronomical Journal] {10.3847/1538-3881/ac3746}, 163, 35

\bibitem[\protect\citeauthoryear{Burrows \& Sharp}{Burrows \&
  Sharp}{1999}]{burrows_1999}
Burrows A.,  Sharp C.~M.,  1999, \mn@doi [The Astrophysical Journal]
  {10.1086/306811}, 512, 843

\bibitem[\protect\citeauthoryear{Carone et~al.,}{Carone
  et~al.}{2020}]{carone_2020}
Carone L.,  et~al., 2020, \mn@doi [Monthly Notices of the Royal Astronomical
  Society] {10.1093/mnras/staa1733}, 496, 3582

\bibitem[\protect\citeauthoryear{Christie et~al.,}{Christie
  et~al.}{2021}]{christie_2021}
Christie D.~A.,  et~al., 2021, \mn@doi [Monthly Notices of the Royal
  Astronomical Society] {10.1093/mnras/stab2027}, 506, 4500

\bibitem[\protect\citeauthoryear{Christie et~al.,}{Christie
  et~al.}{2022a}]{christie_2022b}
Christie D.~A.,  et~al., 2022a, \mn@doi [The Planetary Science Journal]
  {10.3847/PSJ/ac9dfe}, 3, 261

\bibitem[\protect\citeauthoryear{Christie, Mayne, Gillard, Manners, Hébrard,
  Lines  \& Kohary}{Christie et~al.}{2022b}]{christie_2022}
Christie D.~A.,  Mayne N.~J.,  Gillard R.~M.,  Manners J.,  Hébrard E.,  Lines
  S.,   Kohary K.,  2022b, \mn@doi [Monthly Notices of the Royal Astronomical
  Society] {10.1093/mnras/stac2763}, 517, 1407

\bibitem[\protect\citeauthoryear{Deitrick, Mendonça, Schroffenegger, Grimm,
  Tsai  \& Heng}{Deitrick et~al.}{2020}]{deitrick_thor_2020}
Deitrick R.,  Mendonça J.~M.,  Schroffenegger U.,  Grimm S.~L.,  Tsai S.-M.,
  Heng K.,  2020, \mn@doi [The Astrophysical Journal Supplement Series]
  {10.3847/1538-4365/ab930e}, 248, 30

\bibitem[\protect\citeauthoryear{Drummond, Tremblin, Baraffe, Amundsen, Mayne,
  Venot  \& Goyal}{Drummond et~al.}{2016}]{drummond_2016}
Drummond B.,  Tremblin P.,  Baraffe I.,  Amundsen D.~S.,  Mayne N.~J.,  Venot
  O.,   Goyal J.,  2016, \mn@doi [Astronomy \& Astrophysics]
  {10.1051/0004-6361/201628799}, 594, A69

\bibitem[\protect\citeauthoryear{Drummond, Mayne, Baraffe, Tremblin, Manners,
  Amundsen, Goyal  \& Acreman}{Drummond et~al.}{2018}]{drummond_2018}
Drummond B.,  Mayne N.~J.,  Baraffe I.,  Tremblin P.,  Manners J.,  Amundsen
  D.~S.,  Goyal J.,   Acreman D.,  2018, \mn@doi [Astronomy \& Astrophysics]
  {10.1051/0004-6361/201732010}, 612, A105

\bibitem[\protect\citeauthoryear{Edwards \& Slingo}{Edwards \&
  Slingo}{1996}]{edwards_1996}
Edwards J.~M.,  Slingo A.,  1996, \mn@doi [Quarterly Journal of the Royal
  Meteorological Society] {10.1002/qj.49712253107}, 122, 689

\bibitem[\protect\citeauthoryear{Fauchez et~al.,}{Fauchez
  et~al.}{2022}]{fauchez_2022}
Fauchez T.~J.,  et~al., 2022, \mn@doi [The Planetary Science Journal]
  {10.3847/PSJ/ac6cf1}, 3, 213

\bibitem[\protect\citeauthoryear{Fromang, Leconte  \& Heng}{Fromang
  et~al.}{2016}]{fromang_2016}
Fromang S.,  Leconte J.,   Heng K.,  2016, \mn@doi [Astronomy \& Astrophysics]
  {10.1051/0004-6361/201527600}, 591, A144

\bibitem[\protect\citeauthoryear{Hammond \& Abbot}{Hammond \&
  Abbot}{2022}]{hammond_2022}
Hammond M.,  Abbot D.~S.,  2022, \mn@doi [Monthly Notices of the Royal
  Astronomical Society] {10.1093/mnras/stac228}, 511, 2313

\bibitem[\protect\citeauthoryear{Hardiman, Andrews, White, Butchart  \&
  Edmond}{Hardiman et~al.}{2010}]{hardiman_2010}
Hardiman S.~C.,  Andrews D.~G.,  White A.~A.,  Butchart N.,   Edmond I.,  2010,
  \mn@doi [Journal of the Atmospheric Sciences] {10.1175/2010JAS3355.1}, 67,
  1983

\bibitem[\protect\citeauthoryear{Harris et~al.,}{Harris
  et~al.}{2020}]{harris_2020}
Harris C.~R.,  et~al., 2020, \mn@doi [Nature] {10.1038/s41586-020-2649-2}, 585,
  357

\bibitem[\protect\citeauthoryear{Hattersley et~al.,}{Hattersley
  et~al.}{2023}]{hattersley_2023}
Hattersley R.,  et~al., 2023, \mn@doi [Zenodo] {10.5281/zenodo.8305232}

\bibitem[\protect\citeauthoryear{Heng, Menou  \& Phillipps}{Heng
  et~al.}{2011}]{heng_2011}
Heng K.,  Menou K.,   Phillipps P.~J.,  2011, \mn@doi [Monthly Notices of the
  Royal Astronomical Society] {10.1111/j.1365-2966.2011.18315.x}, 413, 2380

\bibitem[\protect\citeauthoryear{Hunter}{Hunter}{2007}]{hunter_2007}
Hunter J.~D.,  2007, \mn@doi [Computing in Science \& Engineering]
  {10.1109/MCSE.2007.55}, 9, 90

\bibitem[\protect\citeauthoryear{Koll \& Komacek}{Koll \&
  Komacek}{2018}]{koll_2018}
Koll D. D.~B.,  Komacek T.~D.,  2018, \mn@doi [The Astrophysical Journal]
  {10.3847/1538-4357/aaa3de}, 853, 133

\bibitem[\protect\citeauthoryear{Komacek, Showman  \& Parmentier}{Komacek
  et~al.}{2019}]{komacek_2019}
Komacek T.~D.,  Showman A.~P.,   Parmentier V.,  2019, \mn@doi [The
  Astrophysical Journal] {10.3847/1538-4357/ab338b}, 881, 152

\bibitem[\protect\citeauthoryear{Li \& Goodman}{Li \& Goodman}{2010}]{li_2010}
Li J.,  Goodman J.,  2010, \mn@doi [The Astrophysical Journal]
  {10.1088/0004-637X/725/1/1146}, 725, 1146

\bibitem[\protect\citeauthoryear{Lines et~al.,}{Lines
  et~al.}{2018}]{lines_2018a}
Lines S.,  et~al., 2018, \mn@doi [Astronomy \& Astrophysics]
  {10.1051/0004-6361/201732278}, 615, A97

\bibitem[\protect\citeauthoryear{Liu \& Showman}{Liu \&
  Showman}{2013}]{liu_2013}
Liu B.,  Showman A.~P.,  2013, \mn@doi [The Astrophysical Journal]
  {10.1088/0004-637X/770/1/42}, 770, 42

\bibitem[\protect\citeauthoryear{Lodders, Palme  \& Gail}{Lodders
  et~al.}{2009}]{lodders_2009}
Lodders K.,  Palme H.,   Gail H.~P.,  2009, \mn@doi [Landolt Börnstein]
  {10.1007/978-3-540-88055-4_34}, 4B, 712

\bibitem[\protect\citeauthoryear{Mayne, Baraffe, Acreman, Smith, Wood,
  Amundsen, Thuburn  \& Jackson}{Mayne et~al.}{2014a}]{mayne_2014b}
Mayne N.~J.,  Baraffe I.,  Acreman D.~M.,  Smith C.,  Wood N.,  Amundsen D.~S.,
   Thuburn J.,   Jackson D.~R.,  2014a, \mn@doi [Geoscientific Model
  Development] {10.5194/gmd-7-3059-2014}, 7, 3059

\bibitem[\protect\citeauthoryear{Mayne et~al.,}{Mayne
  et~al.}{2014b}]{mayne_2014a}
Mayne N.~J.,  et~al., 2014b, \mn@doi [Astronomy \& Astrophysics]
  {10.1051/0004-6361/201322174}, 561, A1

\bibitem[\protect\citeauthoryear{Mayne et~al.,}{Mayne
  et~al.}{2017}]{mayne_2017}
Mayne N.~J.,  et~al., 2017, \mn@doi [Astronomy \& Astrophysics]
  {10.1051/0004-6361/201730465}, 604, A79

\bibitem[\protect\citeauthoryear{Mayne, Drummond, Debras, Jaupart, Manners,
  Boutle, Baraffe  \& Kohary}{Mayne et~al.}{2019}]{mayne_2019}
Mayne N.~J.,  Drummond B.,  Debras F.,  Jaupart E.,  Manners J.,  Boutle I.~A.,
   Baraffe I.,   Kohary K.,  2019, \mn@doi [The Astrophysical Journal]
  {10.3847/1538-4357/aaf6e9}, 871, 56

\bibitem[\protect\citeauthoryear{Menou}{Menou}{2012}]{menou_2012}
Menou K.,  2012, \mn@doi [The Astrophysical Journal]
  {10.1088/2041-8205/744/1/L16}, 744, L16

\bibitem[\protect\citeauthoryear{Parmentier, Showman  \& Lian}{Parmentier
  et~al.}{2013}]{parmentier_2013}
Parmentier V.,  Showman A.~P.,   Lian Y.,  2013, \mn@doi [Astronomy \&
  Astrophysics] {10.1051/0004-6361/201321132}, 558, A91

\bibitem[\protect\citeauthoryear{Perna, Menou  \& Rauscher}{Perna
  et~al.}{2010}]{perna_2010a}
Perna R.,  Menou K.,   Rauscher E.,  2010, \mn@doi [The Astrophysical Journal]
  {10.1088/0004-637X/719/2/1421}, 719, 1421

\bibitem[\protect\citeauthoryear{Polichtchouk, Cho, Watkins, Thrastarson,
  Umurhan  \& de~la Torre~Juárez}{Polichtchouk
  et~al.}{2014}]{polichtchouk_2014}
Polichtchouk I.,  Cho J.-K.,  Watkins C.,  Thrastarson H.,  Umurhan O.,   de~la
  Torre~Juárez M.,  2014, \mn@doi [Icarus] {10.1016/j.icarus.2013.11.027},
  229, 355

\bibitem[\protect\citeauthoryear{Powell, Louden, Kreidberg, Zhang, Gao  \&
  Parmentier}{Powell et~al.}{2019}]{powell_2019}
Powell D.,  Louden T.,  Kreidberg L.,  Zhang X.,  Gao P.,   Parmentier V.,
  2019, \mn@doi [The Astrophysical Journal] {10.3847/1538-4357/ab55d9}, 887,
  170

\bibitem[\protect\citeauthoryear{Rauscher \& Menou}{Rauscher \&
  Menou}{2013}]{rauscher_2013}
Rauscher E.,  Menou K.,  2013, \mn@doi [The Astrophysical Journal]
  {10.1088/0004-637X/764/1/103}, 764, 103

\bibitem[\protect\citeauthoryear{Sainsbury-Martinez et~al.,}{Sainsbury-Martinez
  et~al.}{2019}]{sainsbury-martinez_2019}
Sainsbury-Martinez F.,  et~al., 2019, \mn@doi [Astronomy \& Astrophysics]
  {10.1051/0004-6361/201936445}, 632, A114

\bibitem[\protect\citeauthoryear{Schneider, Carone, Decin, Jørgensen  \&
  Helling}{Schneider et~al.}{2022}]{schneider_2022}
Schneider A.~D.,  Carone L.,  Decin L.,  Jørgensen U.~G.,   Helling C.,  2022,
  \mn@doi [Astronomy \& Astrophysics] {10.1051/0004-6361/202244797}, 666, L11

\bibitem[\protect\citeauthoryear{Sergeev et~al.,}{Sergeev
  et~al.}{2022}]{sergeev_2022}
Sergeev D.~E.,  et~al., 2022, \mn@doi [The Planetary Science Journal]
  {10.3847/PSJ/ac6cf2}, 3, 212

\bibitem[\protect\citeauthoryear{Showman \& Polvani}{Showman \&
  Polvani}{2011}]{showman_2011a}
Showman A.~P.,  Polvani L.~M.,  2011, The Astrophysical Journal, 738, 71

\bibitem[\protect\citeauthoryear{Showman, Cooper, Fortney  \& Marley}{Showman
  et~al.}{2008}]{showman_2008}
Showman A.~P.,  Cooper C.~S.,  Fortney J.~J.,   Marley M.~S.,  2008, \mn@doi
  [The Astrophysical Journal] {10.1086/589325}, 682, 559

\bibitem[\protect\citeauthoryear{Skinner \& Cho}{Skinner \&
  Cho}{2021}]{skinner_2021}
Skinner J.,  Cho J.-K.,  2021, \mn@doi [Monthly Notices of the Royal
  Astronomical Society] {10.1093/mnras/stab971}, 504, 5172

\bibitem[\protect\citeauthoryear{Skinner, Nättilä  \& Cho}{Skinner
  et~al.}{2023}]{skinner_2023}
Skinner J.~W.,  Nättilä J.,   Cho J. Y.-K.,  2023, \mn@doi [Physical Review
  Letters] {10.1103/PhysRevLett.131.231201}, 131, 231201

\bibitem[\protect\citeauthoryear{Tennyson \& Yurchenko}{Tennyson \&
  Yurchenko}{2012}]{tennyson_2012}
Tennyson J.,  Yurchenko S.~N.,  2012, \mn@doi [Monthly Notices of the Royal
  Astronomical Society] {10.1111/j.1365-2966.2012.21440.x}, 425, 21

\bibitem[\protect\citeauthoryear{Tennyson et~al.,}{Tennyson
  et~al.}{2016}]{tennyson_2016}
Tennyson J.,  et~al., 2016, \mn@doi [Journal of Molecular Spectroscopy]
  {10.1016/j.jms.2016.05.002}, 327, 73

\bibitem[\protect\citeauthoryear{Tremblin, Amundsen, Mourier, Baraffe,
  Chabrier, Drummond, Homeier  \& Venot}{Tremblin et~al.}{2015}]{tremblin_2015}
Tremblin P.,  Amundsen D.~S.,  Mourier P.,  Baraffe I.,  Chabrier G.,  Drummond
  B.,  Homeier D.,   Venot O.,  2015, \mn@doi [The Astrophysical Journal]
  {10.1088/2041-8205/804/1/L17}, 804, L17

\bibitem[\protect\citeauthoryear{Tremblin et~al.,}{Tremblin
  et~al.}{2017}]{tremblin_2017}
Tremblin P.,  et~al., 2017, \mn@doi [The Astrophysical Journal]
  {10.3847/1538-4357/aa6e57}, 841, 30

\bibitem[\protect\citeauthoryear{Tsai, Dobbs-Dixon  \& Gu}{Tsai
  et~al.}{2014}]{tsai_2014}
Tsai S.-M.,  Dobbs-Dixon I.,   Gu P.-G.,  2014, \mn@doi [The Astrophysical
  Journal] {10.1088/0004-637X/793/2/141}, 793, 141

\bibitem[\protect\citeauthoryear{Tsai, Moses, Powell  \& Lee}{Tsai
  et~al.}{2023}]{tsai_2023}
Tsai S.-M.,  Moses J.~I.,  Powell D.,   Lee E. K.~H.,  2023, \mn@doi [The
  Astrophysical Journal Letters] {10.3847/2041-8213/ad1405}, 959, L30

\bibitem[\protect\citeauthoryear{Tsai, Innes, Wogan  \& Schwieterman}{Tsai
  et~al.}{2024}]{tsai_2024b}
Tsai S.-M.,  Innes H.,  Wogan N.~F.,   Schwieterman E.~W.,  2024, \mn@doi [The
  Astrophysical Journal Letters] {10.3847/2041-8213/ad3801}, 966, L24

\bibitem[\protect\citeauthoryear{Turbet et~al.,}{Turbet
  et~al.}{2022}]{turbet_2022}
Turbet M.,  et~al., 2022, \mn@doi [The Planetary Science Journal]
  {10.3847/PSJ/ac6cf0}, 3, 211

\bibitem[\protect\citeauthoryear{Watkins \& Cho}{Watkins \&
  Cho}{2010}]{watkins_2010}
Watkins C.,  Cho J. Y.-K.,  2010, \mn@doi [The Astrophysical Journal]
  {10.1088/0004-637X/714/1/904}, 714, 904

\bibitem[\protect\citeauthoryear{Wood et~al.,}{Wood et~al.}{2014}]{wood_2014}
Wood N.,  et~al., 2014, \mn@doi [Quarterly Journal of the Royal Meteorological
  Society] {10.1002/qj.2235}, 140, 1505

\bibitem[\protect\citeauthoryear{Zamyatina et~al.,}{Zamyatina
  et~al.}{2023}]{zamyatina_2023}
Zamyatina M.,  et~al., 2023, \mn@doi [Monthly Notices of the Royal Astronomical
  Society] {10.1093/mnras/stac3432}, 519, 3129

\bibitem[\protect\citeauthoryear{Zamyatina et~al.,}{Zamyatina
  et~al.}{2024}]{zamyatina_2024}
Zamyatina M.,  et~al., 2024, \mn@doi [Monthly Notices of the Royal Astronomical
  Society] {10.1093/mnras/stae600}, 529, 1776

\makeatother
\end{thebibliography}



\appendix

\section{Eddy Momentum Fluxes}
\label{Apdx:Eddy}
As the equatorial jet is driven by vertical and meridional eddy momentum fluxes \citep{showman_2011a}, we include these plots of the eddy momentum flux gradients for interested readers.  Generally, with the exception of the unphysical cases where a counter-rotating jet forms, the flux gradients show the same qualitative morphology for all cases, which is also similar to results found for the radiative cases in the hot Jupiter simulations performed in \citet{mayne_2017} (cf., their Figure 14, right column).  The decreases in dissipation and damping do result in changes in the small-scale structure, but we do not investigate the issue any further.   

To compute the relevant flux gradients, we follow \citet{hardiman_2010} and \citet{mayne_2017} in decomposing the flow into the sum of a zonally-averaged component and a perturbation from that mean (e.g., the zonal velocity $u$ becomes $u = \overline{u} + u'$ where $\overline{u}$ and $u'$ are the zonal mean and perturbations, respectively).  The time evolution of the zonally-averaged momentum $\overline{\rho}\,\overline{u}$ can then be written as

\begin{align}
    (\overline{\rho}\,\overline{u})_{,t} = &  -(\overline{\rho'u'})_{,t} + \underbrace{\overline{\rho G_\lambda}}_\text{Zonal Forcing}  \nonumber \\
    & + \underbrace{2\Omega\overline{\rho v}\sin\phi  - 2\Omega\overline{\rho w}\cos\phi}_\text{Coriolis} \nonumber \\
    &  \underbrace{-\frac{[\overline{(\rho v)'u'}\cos^2\phi]_{,\phi}}{r\cos^2\phi}}_\text{Merid. Eddy Transport} \,\, \underbrace{-\frac{[\overline{(\rho w)'u'}r^3]_{,r}}{r^3}
    \vphantom{\frac{[\overline{(\rho v)'u'}\cos^2\phi]_{,\phi}}{r\cos^2\phi}}}_\text{Vert. Eddy Transport} \nonumber \\
    & \underbrace{-\frac{(\overline{\rho v}\,\overline{u}\cos^2\phi)_{,\phi}}{r\cos^2\phi}}_\text{Mean Merid. Transport}\,\,  \underbrace{-\frac{(\overline{\rho w}\,\overline{u}r^3)_{,r}}{r^3} \vphantom{-\frac{(\overline{\rho v}\,\overline{u}\cos^2\phi)_{,\phi}}{r\cos^2\phi}} }_\text{Mean Vert. Transport}
    \label{Eqn:MomTerms}
\end{align}

\noindent where various transport terms are labelled therein.

\begin{figure*}
	\includegraphics[]{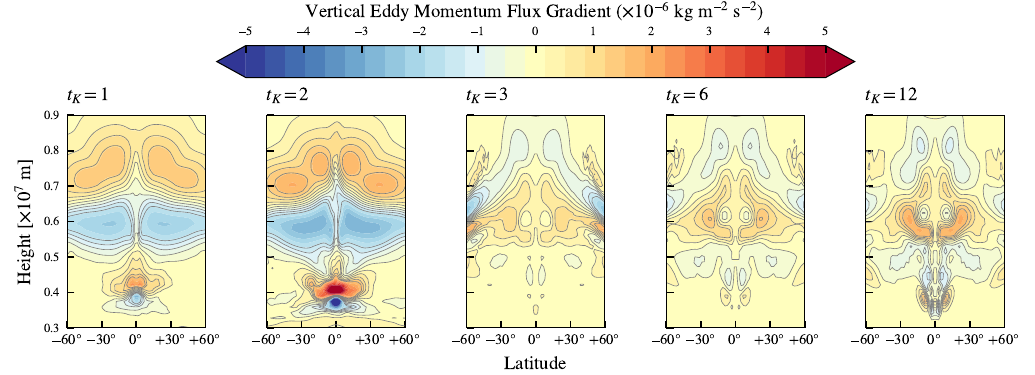}
    \caption{The vertical eddy momentum flux gradient for the average atmospheric state between days 900 and 1000 for each value of $t_K$ in the $\eta_s=0.75$ parameter study (see Equation \ref{Eqn:MomTerms}). For the two over-damped cases, $t_K=1$ and $2$, the vertical transport of eddy momentum can be seen to be vastly different from the cases in which a super-rotating equatorial jet develops, $t_K=3,6,$ and $12$. Note that the vertical coordinate in these plots is the height $z$, as opposed to the pressure as used in other plots and that the plots show only the central region of the computational domain.}
    \label{Fig:eddymom_vert}
\end{figure*}

\begin{figure*}
	\includegraphics[]{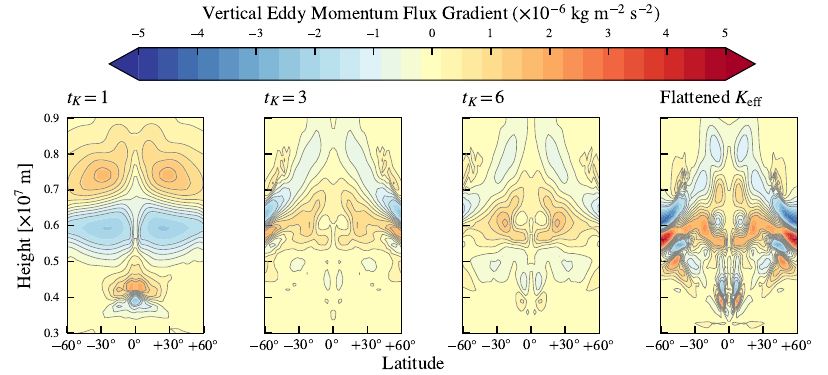}
    \caption{The vertical eddy momentum flux gradient (see Equation \ref{Eqn:MomTerms}) for the average atmospheric state between days 900 and 1000 for the $\eta_s=0.9$ cases (three leftmost panels) and the case with a flattened $K_\mathrm{eff}$ and $\eta_s=0.75$ (rightmost panel). Note that the vertical coordinate in these plots is the height $z$, as opposed to the pressure as used in other plots and that the plots show only the central region of the computational domain.}
    \label{Fig:eddymom_vert_eta9}
\end{figure*}

\begin{figure*}
	\includegraphics[]{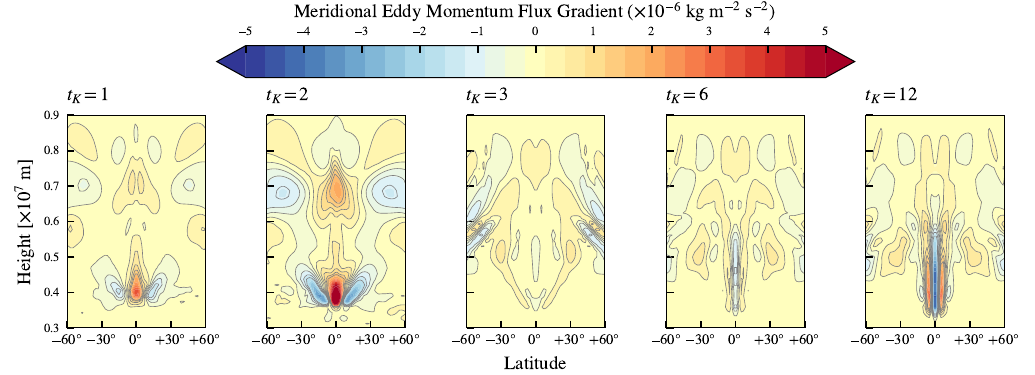}
    \caption{The meridional eddy momentum flux gradient for the average atmospheric state between days 900 and 1000 for each value of $t_K$ in the $\eta_s=0.75$ parameter study (see Equation \ref{Eqn:MomTerms}). For the two over-damped cases, $t_K=1$ and $2$, the vertical transport of eddy momentum can be seen to be vastly different from the cases in which a super-rotating equatorial jet develops, $t_K=3,6,$ and $12$. Note that the vertical coordinate in these plots is the height $z$, as opposed to the pressure as used in other plots and that the plots show only the central region of the computational domain.}
    \label{Fig:eddymom_lat}
\end{figure*}

\begin{figure*}
	\includegraphics[]{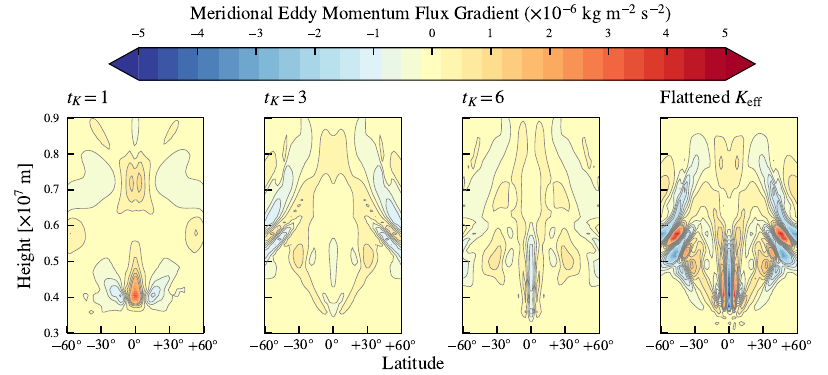}
    \caption{The meridional eddy momentum flux gradient (see Equation \ref{Eqn:MomTerms}) for the average atmospheric state between days 900 and 1000 for the $\eta_s=0.9$ cases (three leftmost panels) and the case with a flattened $K_\mathrm{eff}$ and $\eta_s=0.75$ (rightmost panel). Note that the vertical coordinate in these plots is the height $z$, as opposed to the pressure as used in other plots and that the plots show only the central region of the computational domain.}
    \label{Fig:eddymom_lat_eta9}
\end{figure*}

\begin{figure*}
	\includegraphics[]{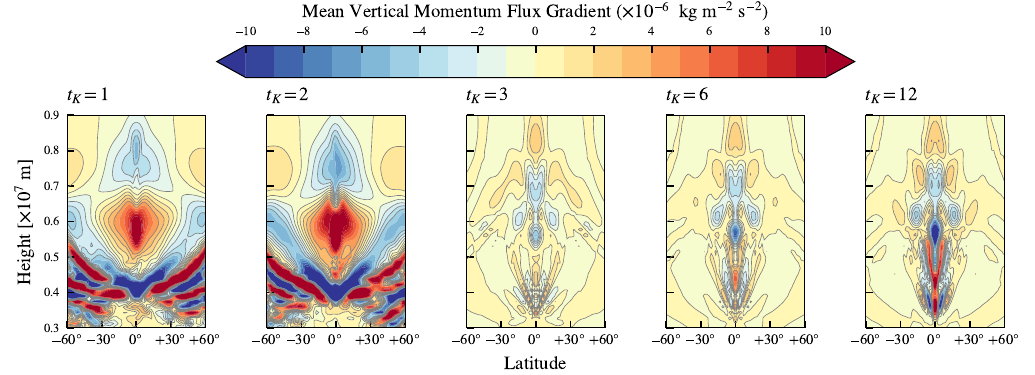}
    \caption{The mean vertical momentum flux gradient for the average atmospheric state between days 900 and 1000 for each value of $t_K$ in the $\eta_s=0.75$ parameter study (see Equation \ref{Eqn:MomTerms}). For the two over-damped cases, $t_K=1$ and $2$, the vertical transport of eddy momentum can be seen to be vastly different from the cases in which a super-rotating equatorial jet develops, $t_K=3,6,$ and $12$. Note that the vertical coordinate in these plots is the height $z$, as opposed to the pressure as used in other plots and that the plots show only the central region of the computational domain.}
    \label{Fig:meanmom_vert_eta75}
\end{figure*}

\begin{figure*}
	\includegraphics[]{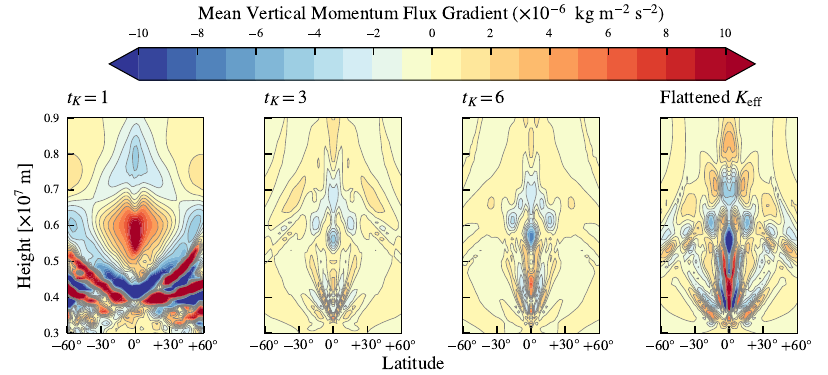}
    \caption{The mean vertical momentum flux gradient (see Equation \ref{Eqn:MomTerms}) for the average atmospheric state between days 900 and 1000 for the $\eta_s=0.9$ cases (three leftmost panels) and the case with a flattened $K_\mathrm{eff}$ and $\eta_s=0.75$ (rightmost panel). Note that the vertical coordinate in these plots is the height $z$, as opposed to the pressure as used in other plots and that the plots show only the central region of the computational domain.}
    \label{Fig:meanmom_vert_eta9}
\end{figure*}

\begin{figure*}
	\includegraphics[]{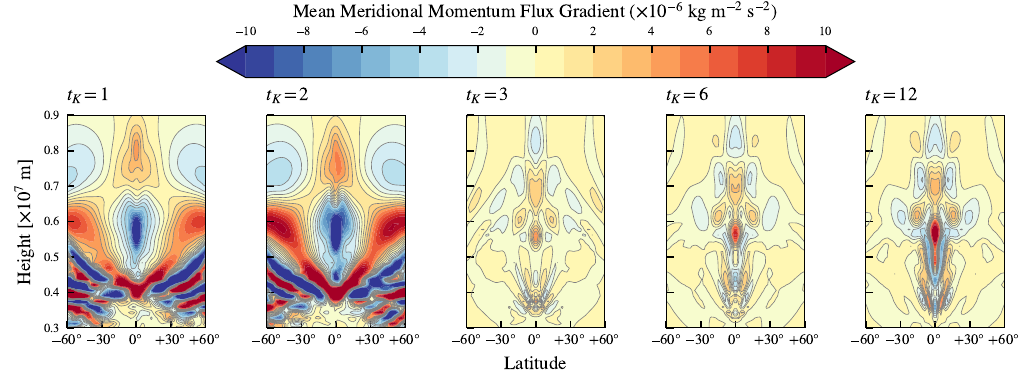}
    \caption{The mean meridional momentum flux gradient for the average atmospheric state between days 900 and 1000 for each value of $t_K$ in the $\eta_s=0.75$ parameter study (see Equation \ref{Eqn:MomTerms}). For the two over-damped cases, $t_K=1$ and $2$, the vertical transport of eddy momentum can be seen to be vastly different from the cases in which a super-rotating equatorial jet develops, $t_K=3,6,$ and $12$. Note that the vertical coordinate in these plots is the height $z$, as opposed to the pressure as used in other plots and that the plots show only the central region of the computational domain.}
    \label{Fig:meanmom_lat_eta75}
\end{figure*}

\begin{figure*}
	\includegraphics[]{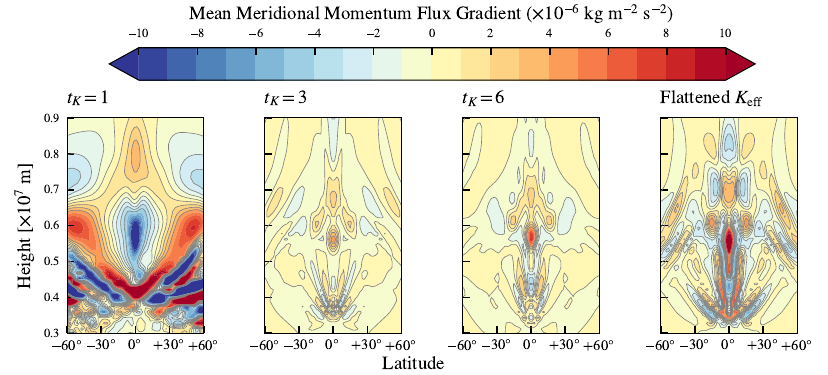}
    \caption{The mean meridional momentum flux gradient (see Equation \ref{Eqn:MomTerms}) for the average atmospheric state between days 900 and 1000 for the $\eta_s=0.9$ cases (three leftmost panels) and the case with a flattened $K_\mathrm{eff}$ and $\eta_s=0.75$ (rightmost panel). Note that the vertical coordinate in these plots is the height $z$, as opposed to the pressure as used in other plots and that the plots show only the central region of the computational domain.}
    \label{Fig:meanmom_lat_eta9}
\end{figure*}

The meridional and vertical eddy momentum flux gradients as well the mean meridional and vertical momentum flux gradients for all the simulations investigated in this paper are shown in  Figures \ref{Fig:eddymom_vert} through \ref{Fig:meanmom_lat_eta9}, inclusive.

\section{Resolution}
\label{Apdx:Res}
 As the effective dissipation coefficient $K_\mathrm{eff}$ scales as $K_\mathrm{eff}\propto \left(\Delta \lambda\right)^2$, a doubling of the longitudinal resolution results in a quartering of $K_\mathrm{eff}$. At the fiducial resolution used in this study, a change from $t_K=1$ to $t_K=3$ -- transitioning from the counter-rotating to the super-rotating cases -- reduces $K_\mathrm{eff}$ by a factor of 2.23.  Thus, one may expect that a $t_K=1$ simulation with the longitudinal resolution doubled would result in a super-rotating jet.   As discussed in \citet{heng_2011}, it is not always possible when investigating the behaviour of a dissipation operator to decouple the effect of the operator from the intrinsic effects of changing the grid.  We nonetheless opt to verify explicitly that the increase in resolution results in the formation of a super-rotating jet.

To this end, we perform a test simulation with $t_K=1$ and the horizontal grid resolution doubled in both the latitude and longitude, resulting in a grid of 288 longitude points and 180 latitude points.  The simulation is able to run with the same dynamical timestep as the fiducial simulations, $\Delta t_\mathrm{dyn}=30\,\mathrm{s}$, avoiding complications in interpreting the results due to the timestep changing as well (see Equation \ref{Eqn:Keff}).  Further refinements in the grid would, unfortunately, require a reduction in $\Delta t_\mathrm{dyn}$, increasing $K_\mathrm{eff}$.  \footnote{While a resolution study would be valuable for the understanding of the behaviour of the {\sc um} generally, especially given the resolution dependence found in the hot Jupiter simulations of \citet{skinner_2021} and the relatively high resolution requirements needed for convergence within their model,  such a study should be designed based on the dissipation requirements to run at the highest resolutions, not as an off-shoot of a study that looks at low-resolution behaviour.} As the intention is to only demonstrate that a jet forms, we run the simulation for 200 days, sufficient for the model to spin-up.

\begin{figure*}
	\includegraphics[]{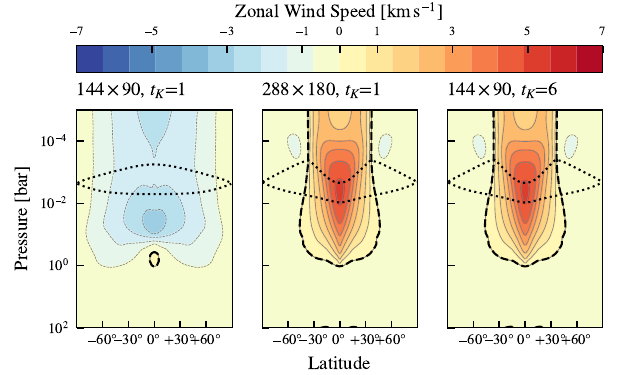}
    \caption{ The mean zonal wind at simulation day 200 for the fiducial simulations with $t_K=1$ (left panel) and $t_K=6$ (right panel) and $\eta_s=0.75$.  The center panel shows the results for a simulation with a doubled horizontal resolution and $t_K=1$.  The dashed line indicates the boundary between super-rotating and counter-rotating flows. The dotted lines indicate the pressure levels at which the pressures partially (lower line) and fully (upper line) intersect the sponge layer. }
    \label{Fig:zonalwind_res}
\end{figure*}

The results from the simulation are shown in Figure \ref{Fig:zonalwind_res}.  As a quartering of $K_\mathrm{eff}$ relative to the $t_K=1$ case is equivalent to increasing $t_K$ to $\sim 5.81$, we compare the results of the higher resolution simulation to both the $t_K=1$ case at the fiducial resolution as well as $t_K=6$ at the fiducial resolution.  As is shown in Figure \ref{Fig:zonalwind_res}, the increased resolution simulation with $t_K=1$ forms a super-rotating jet qualitatively similar to the $t_K=6$ result at the fiducial resolution, with peak mean zonal wind speeds of 5.04 $\mathrm{km\,s^{-1}}$ and 5.01 $\mathrm{km\,s^{-1}}$, respectively, further supporting the idea that the physical breakdown is due to the $K_\mathrm{eff}$.  We see little effect from the resolution in the mean zonal wind or in other quantities (not plotted here) beyond the impact on $K_\mathrm{eff}$.  Resolution dependent effects may become apparent with a reduction in $t_K$ in the increased resolution simulation; however, that is beyond the scope of this paper.

\bsp	
\label{lastpage}
\end{document}